\documentclass[a4paper]{article}

\usepackage[round]{natbib}
\usepackage{amsmath, amssymb, graphicx, amsthm, chngpage, verbatim, hyperref, url, multirow}

\usepackage{amsmath}
\usepackage{amssymb}
\usepackage{amsmath}
\usepackage{amssymb}
\usepackage{amscd}

 \newcommand{\beq}{\begin{equation}}
\newcommand{\eeq}{\end{equation}} 
\newcommand{\bea}{\begin{eqnarray}}
\newcommand{\eea}{\end{eqnarray}}

\newcommand{\Tr}{\mbox{\rm Tr}\,}

  \makeatletter
\newskip\tempskip \def\endproof{{\parfillskip24\p@ plus\@ne
fil\@@par}\tempskip\prevdepth
\ifdim\lastskip=\z@\tempskip\z@\else\vskip-\lastskip
\ifdim\tempskip>4\p@ \tempskip.5\tempskip \else \tempskip\z@\fi\fi
\nobreak\vskip-\baselineskip\vskip-\tempskip\noindent\hbox
to\hsize{\hfill
$\blacksquare$}\par\vskip\tempskip\vskip\abovedisplayskip\@doendpe}
\makeatother \makeatletter
\newskip\tempskip \def\endiproof{{\parfillskip24\p@ plus\@ne
fil\@@par}\tempskip\prevdepth
\ifdim\lastskip=\z@\tempskip\z@\else\vskip-\lastskip
\ifdim\tempskip>4\p@ \tempskip.5\tempskip \else \tempskip\z@\fi\fi
\nobreak\vskip-\baselineskip\vskip-\tempskip\noindent\hbox
to\hsize{\hfill
$\Box$}\par\vskip\tempskip\vskip\abovedisplayskip\@doendpe}
\makeatother 

\newcommand{\lam}{\lambda}

\newcommand{\lt}{\left}
\newcommand{\me}{\middle}
\newcommand{\rt}{\right}
\newcommand{\la}{\langle}
\newcommand{\ra}{\rangle}

\newcommand{\iae}{\overset.=}
\newcommand{\pr}{\mathrm{Pr}}
\newtheorem{theorem}{Theorem}
\newtheorem{lemma}{Lemma}

\begin{document}

\title{An impossibility theorem for Parameter Independent hidden variable theories}

\author{Gijs Leegwater\footnote{Erasmus University Rotterdam, Faculty of Philosophy, Burg. Oudlaan 50, 3062 PA Rotterdam, The Netherlands. E-mail: leegwater@fwb.eur.nl}}

\date{March 31, 2016}

\maketitle
\begin{abstract}
Recently, Roger Colbeck and Renato Renner (C\&R) have claimed that `[n]o extension of quantum theory can have improved predictive power' \citep{Colb2011, Colb2012}. If correct, this is a spectacular impossibility theorem for hidden variable theories, which is more general than the theorems of \citet{Bell1964} and \citet{Legg2003}. Also, C\&R have used their claim in attempt to prove that a system's quantum-mechanical wave function is in a one-to-one correspondence with its `ontic' state \citep{Colb2012b}. C\&R's claim essentially means that in any hidden variable theory that is compatible with quantum-mechanical predictions, probabilities of measurement outcomes are independent of these hidden variables. This makes such variables otiose. On closer inspection, however, the generality and validity of the claim can be contested. First, it is based on an assumption called `Freedom of Choice'. As the name suggests, this assumption involves the independence of an experimenter's choice of measurement settings. But in the way C\&R define this assumption, a no-signalling condition is surreptitiously presupposed, making the assumption less innocent than it sounds. When using this definition, any hidden variable theory violating Parameter Independence, such as Bohmian Mechanics, is immediately shown to be incompatible with quantum-mechanical predictions. Also, the argument of C\&R is hard to follow and their mathematical derivation contains several gaps, some of which cannot be closed in the way they suggest. We shall show that these gaps can be filled. The issue with the `Freedom of Choice' assumption can be circumvented by explicitly assuming Parameter Independence. This makes the result less general, but better founded. We then obtain an impossibility theorem for hidden variable theories satisfying Parameter Independence only. As stated above, such hidden variable theories are impossible in the sense that any supplemental variables have no bearing on outcome probabilities, and are therefore trivial. So, while quantum mechanics itself satisfies Parameter Independence, if a variable is added that changes the outcome probabilities, however slightly, Parameter Independence must be violated.
\end{abstract}

\section{Introduction}
\noindent In 1935, Einstein, Podolsky and Rosen famously argued that quantum mechanics is incomplete and that there might be another theory that does provide a complete description of physical reality \citep{EPR1935}. One class of candidates for such a theory is the class of so-called `hidden variable theories', which supplement the quantum state with extra variables.\footnote{In the literature, sometimes the term `hidden variable theory' is understood to refer only to deterministic theories. Instead, we have a very general use of the term in mind: theories that add, in addition to the quantum state, an extra variable to the description of a system.} Hidden variable theories have indeed been developed, for example the de Broglie--Bohm theory  \citep{Bohm1952}, which is deterministic and complete in Einstein's sense. However, a number of impossibility theorems have been derived, showing that large classes of possible hidden variable theories are incompatible with quantum-mechanical predictions. John Bell proved such incompatibility for local deterministic hidden variable theories \citep{Bell1964}, as well as for local stochastic hidden variable theories \citep{Bell1976}, while the incompatibility of `crypto-nonlocal' theories was proven by \citet{Legg2003}. Still, a large class of hidden variable theories, like the de Broglie--Bohm theory, remains unscathed by these impossibility theorems.

The hidden variable theories shown to be incompatible by Bell are theories satisfying a criterion called Factorizability \citep{Fine1982}, which is equivalent to the conjunction of two locality conditions coined by Abner \citet{Shim1984}: Parameter Independence (ParInd) and Outcome Independence (OutInd). Therefore, any hidden variable theory compatible with quantum mechanics violates at least one of these two conditions. In this article we claim something stronger: any hidden variable theory compatible with quantum mechanics violates ParInd, except for `trivial' hidden variable theories, where the values of the hidden variables have no bearing on measurement outcome probabilities.

This article is based on recent work by Roger Colbeck and Renato Renner (C\&R), who have claimed that they have derived an even more general impossibility theorem \citep{Colb2011,Colb2012}. Stating that `[n]o extension of quantum theory can have improved predictive power', they essentially claim that \emph{any} non-trivial hidden variable theory, also if it violates ParInd (like the de Broglie--Bohm theory), is incompatible with quantum-mechanical predictions. Given the wide scope of this claim, this would be a spectacular result, which would to a great extent put constraints on any possible future theory replacing quantum mechanics.

However, C\&R's claim crucially hinges on an assumption dubbed `Freedom of Choice'. As the name suggests, this assumption is meant to be about the freedom of experimenters to choose their measurement settings. From this assumption, C\&R derive `no-signalling', which is essentially equal to ParInd. Nevertheless, when inspecting the way `Freedom of Choice' is defined, it becomes apparent that ParInd is in fact part of this assumption. Most criticism of C\&R's work focuses on this issue \citep{Ghir2013, Ghir2013b, Colb2013, Leif2014, Land2015}. We agree with the criticism: C\&R's `Freedom of Choice' assumption is much stronger than its name suggests. Therefore, while the impression is given that any hidden variable theory with free experimenters is shown to be incompatible with QM, in fact their result applies to a smaller class of hidden variable theories: those satisfying ParInd. The de Broglie--Bohm theory, which violates ParInd, is therefore not shown to be incompatible after all.

If the above issue was the only problem with C\&R's work, the result of the present article could easily be achieved by adding ParInd as an explicit assumption. The theorem would then still be an interesting impossibility theorem, being more general than the theorems of Bell and Leggett. However, there are more shortcomings in the work of C\&R. First, it is hard to understand, even for experts. Valerio \citet{Scar2013} says:

\begin{quote}
`Beyond the case of the maximally entangled state, which had been settled in a previous paper, they prove something that I honestly have not fully understood. Indeed, so many other colleagues have misunderstood this work, that the authors prepared a page of FAQs [\citep{Colb2010}] (extremely rare for a scientific paper) and a later, clearer version [\citep{Colb2012}].'
\end{quote}

The case of the maximally entangled state that Scarani refers to corresponds to the triviality claim of C\&R restricted to local measurements on a Bell state. This result consists of the statement that not only the quantum-mechanical outcome probabilities, but also the outcome probabilities in any hidden variable theory equal $1/2$ (in the present paper, this result is presented in Section \ref{SecBell}). Some authors, for example Antonio \citet{DiLo2012}, appear to have understood C\&R as deriving only this result, which, as Scarani alludes to, had been derived before. Actually, for C\&R this is only the first step in proving the more general theorem that probabilities in hidden variable theories are always equal to the quantum-mechanical probabilities.

More importantly, C\&R's derivation contains gaps, of which some are allegedly filled in other publications, while others remain. One example is their careless handling of limits: in more than one occasion results are derived that only hold approximately, which are then used as if they hold exactly.\footnote{Section \ref{SecDiscussion} contains a more detailed treatment of these gaps.}

Because of these shortcomings, at present no acceptable deduction of the impossibility theorem for hidden variable theories satisfying ParInd exists in the literature. In this article we attempt to repair the shortcomings of C\&R's derivation in order to establish such a deduction. A step that is not explicitly mentioned by C\&R, involving the relation between measurements on entangled states and measurements on non-entangled states, is formulated explicitly. Furthermore, we give a deduction that is mathematically acceptable. We emphasize that this does not consist of simply filling some gaps. For some parts of the deduction to succeed, an entirely new strategy has to be constructed, or so we claim. This is especially the case when taking proper care of all the limits used in the proof. Also, some parts of the deduction can, in our opinion, be considerably simplified, especially the first steps. For these reasons, in this article we do not merely point out all the shortcomings in the original derivation; rather, we construct a new version of it.

C\&R have also used their claim in an attempt to answer the question whether the quantum-mechanical wave function is `ontic' or `epistemic'. Since the appearance of the Pusey-Barret-Rudulph (PBR) theorem \citep{Puse2012}, this is a hotly debated topic. On the basis of their claim, C\&R argue not only that the wave function is ontic, but also that it is in a one-to-one correspondence with its ontic state \citep{Colb2012b}. In the Discussion (Section \ref{SecDiscussion}), we shall consider what remains of this $\psi$-ontology result if C\&R's claim is replaced by the weaker result deduced in this article.

The result will be deduced in several steps. The first steps are quite simple and correspond to results that existed already before the work of C\&R. However, we believe that even for those whom are already familiar with this result, these steps are still of value since they are considerably simplified, only using a triangle inequality and a simple inequality from probability theory. The final steps require more mathematics and may be harder to follow. Most of the mathematics is relegated to appendices, so as not to distract the reader from the central line of reasoning. If the reader want to shorten the reading time, the best section to skip might be Section \ref{SecAny}, because the extent of the generalization (from states with coefficients that are square roots of rational numbers to any coefficients) is relatively small compared to the amount of mathematics needed. It is however a necessary part for deriving the full theoretical result. In the Discussion, I shall mention the most important differences between our deduction and that of C\&R.

\section{Notation}
\noindent Quantum-mechanical systems are referred to by the symbols $A, B, A', B'$ etc. To denote composite systems, the symbols for the subsystems are combined, for example $AB$ and $AA'A''$. The Hilbert space of system $A$ is denoted by $\mathcal{H}_A$, a state as $|\psi\ra_A$ and an operator on $\mathcal{H}_A$ as $U^A$. For notational convenience, the subscript attached to a state may be omitted when no confusion is possible, especially when large composite systems like $AA'A''BB'B''$ are involved. The symbol $\otimes$ for taking tensor products is also often omitted, and we freely change the order of states when combining systems, so that we can write
\begin{align}
U^{AB}\lt(|i\ra_A \otimes |j\ra_B\rt) \equiv U^{AB}\lt(|j\ra_B \otimes |i\ra_A\rt) \equiv U^{AB}\lt(|j\ra_B |i\ra_A\rt).
\end{align}
We also write $[\psi]^A := |\psi\ra_{AA}\la\psi|$, and $\mathbb{N}_r$ for the set $\{0, 1, \dots, r-1\}$. The cardinality of a set $J$ is written as $\#J$, and sequences are notated as $(x_n)_{n=0}^{10}$.

In this article, mainly projective measurements with a finite number of outcomes are considered. Such measurements are defined by a complete set of orthogonal projectors $\{\hat{E}_i^A\}_{i=0}^{d-1}$ with $d \in \mathbb{N}$, each projector corresponding to a possible outcome. If all projectors are 1-dimensional, the set of projectors can be written as $\{[i]^A\}_{i=0}^{d-1}$, in which case it is said that the measurement is performed `in the basis $\{|i\ra_A\}_{i=0}^{d-1}$'. We also allow ourselves to say this when $\{|i\ra_A\}_{i=0}^{d-1}$ is not a basis of $\mathcal{H}_A$, but of a strict subspace of $\mathcal{H}_A$. In this case the corresponding complete set of orthogonal projectors is ${\{[i]^A\}_{i=0}^{d-1} \cup \{\mathbb{I}^A - \sum_{i=0}^{d-1} [i]^A\}}$. Equivalently, a projective measurement can be characterized by an observable (a Hermitian operator) of which the eigenspaces corresponding to the eigenvalues equal the ranges of the projectors $\{\hat{E}_i^A\}_{i=0}^{d-1}$. Two observables which are equal up to their eigenvalues represent the same measurement. So, if a measurement is characterized by the complete set of orthogonal projectors $\{\hat{E}_i^A\}_{i=0}^{d-1}$, a corresponding observable is
\begin{align}
\hat{O}^A = \sum_{i=0}^{d-1} e_i \hat{E}^A_i,
\end{align}
where the $e_i \in \mathbb{R}$ are (distinct) eigenvalues. Probabilities of outcomes can be expressed using the projectors:
\begin{align}
\pr^{|\psi\ra_A}\lt(\hat{E}_i^A\rt) = \la\psi|_A \hat{E}_i^A |\psi\ra_A.
\end{align}
The superscript including the state of the system may be omitted if no confusion is possible. Also, sometimes the superscript on the observable is omitted, for example when a general form of an observable is defined which can be applied to multiple systems.

Often measurements are considered on subsystems, and the pure state of the composite system is specified:
\begin{align}
\pr^{|\phi\ra_{AB}}\lt(\hat{E}_i^A\rt).
\end{align}

To express probabilities using observables, we associate to any observable $\hat{O}^A$ a random variable $O^A$. We want to emphasize that we only consider joint distributions of random variables if their associated observables are jointly measurable (and therefore commute). It is well known that if one defines joint distributions for non-commuting observables, then the corresponding random variables are subject to additional constraints, in the form of a Bell inequality \citep{Fine1982}, which we want to avoid. With this in place, probabilities can be expressed using the random variables associated with observables:
\begin{align}
\pr\lt(O^A = e_i\rt) = \pr\lt(\hat{E}_i^A\rt).
\end{align}
We will see in the next section that the hidden variable theories we consider lead to a decomposition of the quantum probabilities. That is, an extra variable $\lam$ is added to each quantum probability, and there is a measure $\mu(\lam)$ such that, when averaging over $\lam$ using this measure, the quantum probability is retrieved. The probabilities in hidden variable theories and in a decomposition are called $\lam$-probabilities. In decompositions, these probabilities are denoted as the quantum probabilities, with $\lam$ added as a subscript:
\begin{align}
\pr_\lam\lt(O^A = e_i\rt) = \pr_\lam\lt(\hat{E}_i^A\rt)_{\hat{O}^A}.\label{LamProb}
\end{align}
In the notation using projectors, we have added an extra subscript indicating the measured observable. This is because a $\lam$-probability might be \emph{contextual}, depending not only on the projector $\hat{E}_i^A$, but also on the other projectors characterizing the measurement.\footnote{Note that this is a specific type of contextuality which is, for example, different from the contextuality considered in the Kochen--Specken Theorem \citep{Koch1975}, which concerns a dependence of \emph{values}, instead of probabilities, on the measurement context.} %

Identities involving $\lam$-probabilities appearing in this article mostly hold `almost everywhere', i.e. for all $\lam$ in a subset $\Omega \subset \Lambda$ with $\mu(\Omega) = 1$, where $\mu$ is a measure on the measurable space $\Lambda$. Wherever this is the case, the symbol $\iae$ is used. Often we will consider $\lam$-probabilities, expressed using a projector, that are almost everywhere independent of the observable that is measured. For example, we might have\footnote{The quantifier in \eqref{NonContextEx} ranges over all observables that include $\hat{E}^A_i$ in their sets of corresponding projectors}
\begin{align}
\forall \hat{O}^A: \pr_\lam \lt( \hat{E}^A_i \rt)_{\hat{O}^A} \iae \frac12.\label{NonContextEx}
\end{align}
In such cases, we allow ourselves to drop the observable from the notation, so the above can be rewritten as
\begin{align}
\pr_\lam \lt( \hat{E}^A_i \rt) \iae \frac12,
\end{align}
although $\pr_\lam \lt( \hat{E}^A_i \rt)$ might not be well-defined for a measure zero subset of $\lam$'s.

A $\lam$-probability is called trivial if it equals the corresponding quantum-mechanical probability for almost every $\lam$. A decomposition is called trivial if all the $\lam$-probabilities occurring in it are trivial. Also, a hidden variable theory is called trivial if all the $\lam$-probabilities occurring in it are trivial.

From Section \ref{SecSqrRat} onwards, limits are taken involving multiple variables. These are always \emph{repeated} limits. If, for example, we write
\begin{align}
\lim_{l \to \infty} \lim_{n \to \infty} f(n,l) = \lim_{l \to \infty} \lt( \lim_{n \to \infty} f(n,l) \rt),
\end{align}
this means that first the limit $n \to \infty$ is taken, and $\lim_{n \to \infty} f(n,l)$ exists. Note that the order of the limits is important: it might be that in the above case $\lim_{l \to \infty} f(n,l)$ does not exist in which case $\lim_{n \to \infty} \lim_{l \to \infty} f(n,l)$  is not well-defined. If $\epsilon > 0$ and $\lim_{l \to \infty} \lim_{n \to \infty} f(n,l) = 0$ then, to find an $n$ and an $l$ such that $|f(n,l)| < \epsilon$, \emph{first} an $l$ large enough must be chosen, and \emph{then} an $n$ large enough.

\section{Theorem}\label{SecTheorem}
\noindent The main result of this article concerns $\lam$-probabilities for measurements on a system $A$ in the state $|\psi\ra_A$, to which a hidden variable theory assigns an extra variable $\lam$.
This $\lam$ is distributed according to some measure $\mu(\lam)$, which might depend not only on the quantum state $|\psi\ra_A$ but also on other factors, like the specific method that was used to prepare the state. Now, given a value of $\lam$, the outcome probabilities of a measurement of system $A$ in the state $|\psi\ra_A$ may differ from the quantum-mechanical probabilities. Also, the $\lam$-probabilities might have additional dependencies compared with quantum probabilities. For example, $\lam$-probabilities may be contextual, as mentioned in the previous section. They may also depend on the specific implementation of the measurement. If a $\lam$-probability for a certain implementation of the measurement were non-trivial, this would generate a non-trivial decomposition. Therefore, it is enough to show that there is no non-trivial decomposition, and additional factors such as the specific implementation of the measurement do not have to be included when writing down $\lam$-probabilities: $\pr_\lam^{|\psi\ra_A} \lt( \hat{E}^A_i \rt)_{\hat{O}^A}$.\footnote{More specifically, factors such as the specific implementation can be omitted because these can be held fixed throughout the derivation, by picking one implementation for every observable that is considered. However, the measurement context (i.e. the observable to which the projector belongs) does have to be explicitly mentioned initially, because this cannot be held fixed throughout the derivation: different contexts are sometimes considered for the same projector. For example, \eqref{SomeFootnoteEq1} is a result for a specific context, but using perfect correlation, this leads to \eqref{SomeFootnoteEq2}, which holds in any context. Thanks go to Guido Bacciagaluppi for pointing out this method.}

Now, given a value of $\lam$ not only the outcome probabilities of a measurement of system $A$ in the state $|\psi\ra_A$ may differ, but also outcome probabilities of measurements on systems which have interacted with $A$. For example, $A$ may be coupled to a system $B$ in the following way:
\begin{align}
|\psi\ra_A|0\ra_B \mapsto \sum_{j=0}^{d-1} \hat{E}^A_j |\psi\ra_A |j\ra_B
\end{align}
where $\{\hat{E}^A_j\}_{j=0}^{d-1}$ is a complete set of orthogonal projectors and $\{|j\ra_B\}_{j=0}^{d-1}$ a set of orthonormal vectors. Then, a measurement in the basis $\{|j\ra_B\}_{j=0}^{d-1}$ on $B$ can be considered, and for this measurement the $\lam$-probabilities might also differ from the quantum-mechanical probabilities. Like above, there might be additional dependencies, such as on the specific implementation of the interaction between $A$ and $B$. Again, a non-trivial $\lam$-probability would imply a non-trivial decomposition, and therefore these additional dependencies can be      omitted, so we write down the $\lam$-probability as
\begin{align}
\pr_\lam^{|\phi\ra_{AB}} \lt( [i]^B \rt)_{\hat{O}^B}.
\end{align}
So, here $\lam$ still refers to the variable assigned to system $A$ when it was in the state $|\psi\ra_A$. We will also consider measurements on more complicated composite systems like $AA'A''BB'B''$, and also in these cases, $\lam$ always refers to the original system $A$, and there is only one measure $\mu(\lam)$ that is considered.

We can now further impose conditions on such decompositions. For example, we might impose \emph{non-contextuality}, meaning that the $\lam$-probability of a measurement outcome only depends on the corresponding projector, and not on the other projectors characterizing the measurement. This is however a strong condition, as it follows from Gleason's theorem that such decompositions are trivial if the dimension of the Hilbert space of the system is at least 3.\footnote{According to Gleason's theorem \citep{Glea1957}, for any Hilbert space with dimension at least 3, any probability measure over the projectors corresponds to a unique density matrix $\rho$ satisfying $\pr \lt( \hat{E} \rt) = \Tr \lt( \rho \hat{E} \rt)$ for all projectors $\hat{E}$. Now, if $\pr \lt( \hat{E} \rt)$ could be decomposed so that $\pr \lt( \hat{E} \rt) = \int \mathrm{d}\mu(\lam) \pr_\lam \lt( \hat{E} \rt)$ then, by Gleason's theorem, to each $\pr_\lam \lt( \hat{E} \rt)$ there would correspond a density matrix $\rho_\lam$ such that $\pr_\lam \lt( \hat{E} \rt) = \Tr \lt( \rho_\lam \hat{E} \rt)$. But then, $\pr \lt( \hat{E} \rt) = \int \mathrm{d}\mu(\lam) \Tr (\rho_\lam \hat{E}) = \Tr \lt( \lt( \int \mathrm{d}\mu(\lam) \rho_\lam \rt) \hat{E} \rt)$, so that $\int \mathrm{d}\mu(\lam) \rho_\lam = \rho$. If $\rho$ is a pure state, then it can, by the definition of a pure state, only be trivially decomposed, which means that $\rho_\lam \iae \rho$, and therefore $\pr_\lam \lt( \hat{E} \rt) \iae \pr \lt( \hat{E} \rt)$.} We will, however, impose the following conditions:

\begin{itemize}
\item \textbf{CompQuant: Compatibility with Quantum-Mechanical Predictions}\\
We consider theories where the quantum state $|\psi\ra_A$ is supplemented with a hidden variable $\lam$, which assumes a value from the measurable space $\Lambda$. It is assumed that the quantum-mechanical probabilities are retrieved when averaging using the measure $\mu(\lam)$. This means that, for any decomposition generated by the hidden variable theory, for every projective measurement involving state $|\phi\ra$, observable $\hat{O}$ and projector $\hat{E}$,
\begin{align}
\int_{\Lambda} \mathrm{d}\mu(\lam) \pr_\lam^{|\phi\ra}\lt(\hat{E}\rt)_{\hat{O}} = \pr^{|\phi\ra}\lt(\hat{E}\rt) = \la\phi|\hat{E}|\phi\ra.
\end{align}
As discussed above, $\lam$ is the hidden variable assigned to system $A$ when it was in the state $|\psi\ra_A$, although the probabilities considered can be for measurements on systems in other states, e.g. systems which have interacted with $A$. It is also assumed that the measure $\mu$ is independent of what measurement is being performed. This is commonly justified by the fact that experimenters are free to choose their measurements, independent of the value of $\lam$.\footnote{Note that this means that we do not consider retrocausal and superdeterministic models.} %
\item \textbf{ParInd: Parameter Independence}\\
When considering joint measurements, $\lam$-probabilities for measurement outcomes on one subsystem are independent of which measurement is performed, and of whether a measurement is performed at all,\footnote{Not performing a measurement on a system can equivalently be described as `performing a measurement of the observable $\hat{\mathbb{I}}$', which always yields the outcome $1$.} on any other subsystem.\footnote{Note that no requirement is imposed on the spatiotemporal relation between the two measurements constituting the joint measurement: they are not necessarily space-like separated.}%
For example, for measurements on subsystems $A$ and $B$ of a composite system $AB$ this means that
\begin{align}
\sum_{j \in J} \pr_\lam^{|\phi\ra_{AB}}\lt(\hat{E}_i^{A}, \hat{E}_j^{B}\rt)_{\hat{O}^A \otimes \hat{O}^B} &= \sum_{k \in K} \pr_\lam^{|\phi\ra_{AB}}\lt(\hat{E}_i^{A}, \hat{E'}_{k}^{B}\rt)_{\hat{O}^A \otimes \hat{O}^B}\notag\\
&=: \pr_\lam^{|\phi\ra_{AB}}\lt(\hat{E}_i^{A}\rt)_{\hat{O}^A},
\end{align}
where $\{\hat{E}_i^{A}\}_{i \in I}$, $\{\hat{E}_j^{B}\}_{j \in J}$ and $\{\hat{E'}_k^{B}\}_{k \in K}$ are complete sets of orthogonal projectors. Note that ParInd makes the last expression well-defined: without ParInd, it could depend on the type of measurement performed on $B$. Note that this definition of ParInd is more general than the one standardly used in the literature, where it usually only plays a role in EPR-type measurements.
\end{itemize}

Note that it is also implicitly assumed that $\lam$-probabilities for measurements on a certain system are independent of the state of other systems, e.g.
\begin{align}
P_\lam^{|\phi\ra_{AB} \otimes |\zeta\ra_{A'B'}}\lt(\hat{E}^{AB}\rt)_{\hat{O}^{AB}} = P_\lam^{|\phi\ra_{AB} \otimes |\chi\ra_{A'B'}}\lt(\hat{E}^{AB}\rt)_{\hat{O}^{AB}} = P_\lam^{|\phi\ra_{AB}}\lt(\hat{E}^{AB}\rt)_{\hat{O}^{AB}}.\label{PE}
\end{align}
This is used at the end of Section \ref{SecSqrRat}. Denying this doesn't seem to make sense: $\lam$-probabilities could then depend on states of arbitrary other systems, and of course there is always an infinitude of other systems of which the states are continually evolving. We now formulate the main theorem:

\begin{theorem}
Let $A$ be a system to which quantum mechanics assigns the state $|\psi\ra_A$. Consider a projective measurement on $A$ with projectors $\{\hat{E}^A_i\}_{i=0}^{d-1}$. Then, any decomposition satisfying CompQuant and ParInd is trivial, i.e.
\begin{align}
\pr_\lam^{|\psi\ra_A}\lt(\hat{E}^A_i\rt) \iae \pr^{|\psi\ra_A}\lt(\hat{E}^A_i\rt).\label{mainthm}
\end{align}
\end{theorem}
In \cite{Colb2011}, where C\&R first present their triviality claim, the supplemental variable, analogous to $\lam$ in this article, is by them dubbed `additional information' and represented by a discrete random variable $Z$, with distribution $\pr(Z)$. Also, they assume that $Z$ is accessible to experimenters, as an output of an operation that has an input represented by the variable $C$. Note that if this the case, an ensemble of systems in identical quantum-mechanical states can be prepared with a distribution that deviates from the distribution $\pr(Z)$, simply by first preparing an ensemble with distribution $\pr(Z)$ and then discarding some of the systems, the selection depending on the value of $Z$. It is then possible that outcomes of measurements performed on such modified ensembles deviate from the quantum-mechanical predictions. However, there is still compatibility with quantum mechanics in the sense that, when averaging using the original distribution $\pr(Z)$, the quantum-mechanical probabilities are recovered. The fact that C\&R consider $Z$ to be accessible also explains why the condition analogous to ParInd is by them called `no-signalling'. If $\lam$ is accessible to experimenters, a violation of ParInd would allow someone in control of one subsystem to send information to someone in control of another subsystem. In their later article \citep{Colb2012b}, C\&R state that $Z$ can alternatively be interpreted as `forever hidden and hence unlearnable in principle'. In the present article, nothing is assumed about the accessibility of $\lam$, in order to stay as general as possible.

In the following sections, we shall derive the theorem step-by-step. In Section \ref{SecBell}, we repeat the derivation of a known result regarding measurements on Bell states. In Section \ref{SecDDim}, the result is generalized to higher-dimensional states. In Section \ref{SecSqrRat}, states will be considered where the Schmidt coefficients are square roots of rational numbers, while in Section \ref{SecAny} the generalization is made to arbitrary entangled states. In Section \ref{SecAnyProj}, the final generalization is made to any projective measurement and to POVM\footnote{POVM stands for Positive Operator-Valued Measure. A detailed treatment of these different types of measurements can be found in \citet{Busc1996}.} measurements, including measurements on systems in non-entangled states. %

The following table gives an overview of the intermediate steps leading to the final result, and the assumptions made at each step.\footnote{$\hat{E}^A$ is a projector, while $\hat{F}^A$ is a positive operator.}

\begin{center}
\begin{tabular}{|c|c|c|} \hline
Section	&State	&Result \\ \hline \hline
4		&$1/\sqrt2\lt(|0\ra_A|0\ra_B + |1\ra_A|1\ra_B\rt)$	&$\pr_\lam\lt([0]^A\rt) \iae \pr\lt([0]^A\rt) = 1/2$ \\ \hline
\multirow{2}{*}{5}		&$\sum_i c_i |i\ra_A |i\ra_B$	&$c_j = c_k \Rightarrow \pr_\lam\lt([j]^A\rt) \iae \pr_\lam\lt([k]^A\rt)$	 \\ \cline{2-3}
		&$\sum_{i=0}^{d-1}1/\sqrt{d}|i\ra_A |i\ra_B$	&$\forall i: \pr_\lam\lt([i]^A\rt) \iae \pr\lt([i]^A\rt) = 1/d$	 \\ \hline
6		&$\sum_i c_i |i\ra_A |i\ra_B; \forall i: c_i^2 \in \mathbb{Q}$	&$\forall i: \pr_\lam\lt([i]^A\rt) \iae \pr\lt([i]^A\rt) = c_i^2$	 \\ \hline
7		&$\sum_i c_i |i\ra_A |i\ra_B; \forall i: c_i \in \mathbb{R}$	&$\forall i: \pr_\lam\lt([i]^A\rt) \iae \pr\lt([i]^A\rt) = c_i^2$	\\ \hline
\multirow{2}{*}{8}	&$|\psi\ra_A$ 	&$\forall \hat{E}^A\!: \pr_\lam\lt(\hat{E}^A\rt) \iae \pr\lt(\hat{E}^A\rt) = \lt\|\hat{E}^A|\psi\ra_A\rt\|^2$	 		  \\ \cline{2-3}
		&$|\psi\ra_A$ 					&$\forall \hat{F}^A\!: \pr_\lam \lt( \hat{F}^A \rt) \iae \pr \lt( \hat{F}^A \rt) = {}_A\la\psi|\hat{F}^A|\psi\ra_A$ \\ \hline
\end{tabular}
\end{center}
\section{Triviality for Bell states}\label{SecBell}
\noindent Consider a bipartite system $AB$ prepared in a Bell state:
\begin{align}
|\phi\ra_{AB} := \frac1{\sqrt2}(|0\ra_A|0\ra_B + |1\ra_A|1\ra_B) \quad \in \mathcal{H}_A \otimes \mathcal{H}_B = \mathbb{C}^2 \otimes \mathbb{C}^2,\label{BellState}
\end{align}
where $\{|0\ra_A, |1\ra_A\}$ and $\{|0\ra_B, |1\ra_B\}$ are orthonormal bases. Consider a measurement on subsystem $A$ in the basis $\{|0\ra_A, |1\ra_A\}$. This measurement has two possible outcomes, each with a quantum-mechanical probability $1/2$. The result of this section will be that in any decomposition satisfying ParInd and CompQuant, the $\lam$-probabilities of these outcomes also equal $1/2$, rendering such a decomposition trivial with respect to this measurement.

Define a continuous set of observables for $A$ and $B$
\begin{align}
\hat{O}_\theta &:= -1 \cdot [\theta] + 1 \cdot [\theta + \pi], \mbox{ where } \notag\\
 |\theta\ra &:= \cos(\theta/2)|0\ra + \sin(\theta/2)|1\ra \; \mbox{ and } \; \theta \in [0, \pi]. \label{DefOtheta}
\end{align}
Note that $\la \theta | \theta + \pi \ra = 0$, and $\hat{O}_\theta$ has eigenvectors $|\theta\ra, |\theta + \pi\ra$ with eigenvalues $-1, +1$. For any $N \in \mathbb{N}$, define the following discrete subset of these observables:
\begin{align}
\hat{A}_{N,a}	&:= \hat{O}^A_{a\pi/2N}, \quad	 a \in	\{0,2,\dots,2N\} =: \mathcal{A}_N;\notag\\
\hat{B}_{N,b}	&:= \hat{O}^B_{b\pi/2N}, \quad	 b \in	\{1,3,\dots,2N-1\} =: \mathcal{B}_N;\notag\\
\hat{A}_0	&:= \hat{A}_{N,0}.\label{AiBjDef}
\end{align}
The last definition is unambiguous because $\hat{A}_{N,0} = \hat{O}^A_0$ is independent of $N$. Note that the observables $\hat{A}_{N,2N}$ and $\hat{A}_0$ correspond to the measurement in the basis $\{|0\ra_A, |1\ra_A\}$. Also note that $\hat{A}_{N,2N} = -\hat{A}_0$, which means that $A_{N,2N} = -A_0$. Define a correlation measure\footnote{In all summations involving $a$ and $b$, it is assumed that $a \in \mathcal{A}_N$ and  $b \in \mathcal{B}_N$.}
\begin{align}
I_N := \sum_{|a-b|=1} \pr^{|\phi\ra_{AB}}(A_{N,a} \ne B_{N,b}).\label{DefIN}
\end{align}
\begin{figure}
\begin{center}
\includegraphics[width=12cm]{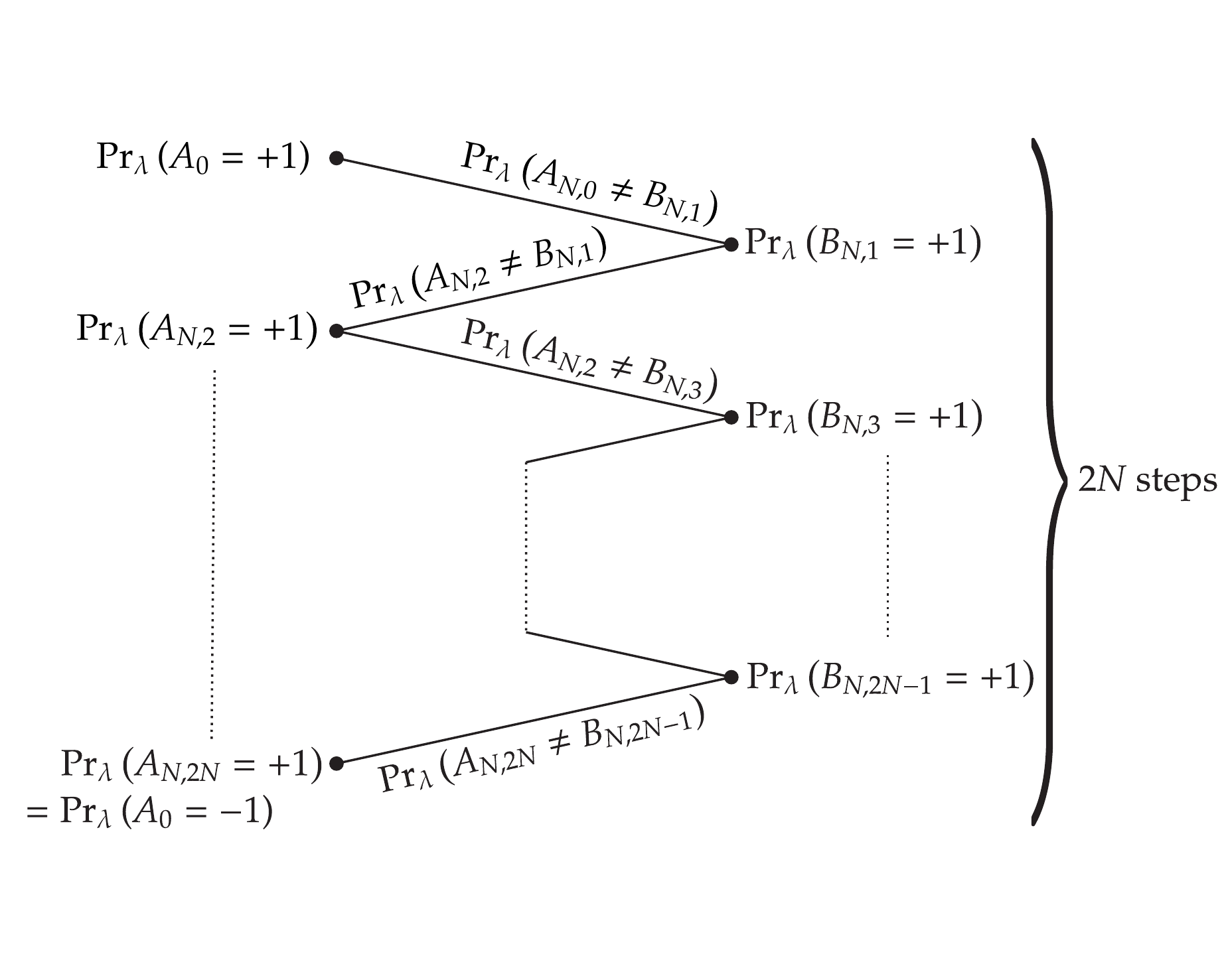}
\caption{The probabilities displayed at two subsequent points differ no more than the probability displayed at the line between these two points. Therefore, the difference between $\pr_\lam( A_0 = +1 )$ and $\pr_\lam( A_0 = -1 )$ is bounded above by $I_N$, which is the sum of $2N$ probabilities of the form $\pr_\lam (A_{N,a} \ne B_{N,b})$ with $|a-b|=1$. The $\lam$-average of each of these probabilities is smaller than $\pi^2 / (16N^2)$. Therefore, the $\lam$-averaged difference between $\pr_\lam( A_0 = +1 )$ and $\pr_\lam( A_0 = -1 )$ is bounded above by $2N \pi^2 / (16N^2) = \pi^2/(8N)$. Since this holds for all $N \in \mathbb{N}$, the $\lam$-averaged difference equals zero, and therefore $\pr_\lam( A_0 = +1 ) = \pr_\lam( A_0 = -1 ) = 1/2$ for almost all $\lam$.}\label{figCBI}
\end{center}
\end{figure}

Before turning to the formal proof, the following should provide an intuition of the idea behind it. The correlation measure above is inspired by the use of \emph{chained Bell inequalities}, which were introduced in \cite{Pear1970} and \cite{Brau1989}. An important ingredient of chained Bell inequalities is that correlations are considered between outcomes when jointly measuring $\hat{O}^A_\theta$ and $\hat{O}^B_{\theta'}$, with a small difference between $\theta$ and $\theta'$. The correlation when jointly measuring two consecutive observables from the list $\hat{A}_0, \hat{B}_{N,1}, \dots, \hat{B}_{N,2N-1}, \hat{A}_{N,2N}$ gets stronger as $N$ gets larger: $\pr(A_{N,a} \ne B_{N,b})$ gets closer to $0$, while the difference between $\theta$ and $\theta'$ becomes smaller. Combining all the correlations between outcomes of consecutive observables, which is what happens in the definition of $I_N$ in \eqref{DefIN}, an upper bound can be derived for the difference in $\lam$-probabilities of the outcomes for the first and the last observables, i.e. $\hat{A}_0$ and $\hat{A}_{N,2N} = -\hat{A}_0$. It turns out that as $N$ gets larger, the correlations become so much stronger that the $\lam$-probabilities for the values of $A_0$ and $-A_0$ must be equal, i.e. they must both equal $1/2$. This is related to the fact that as $N \to \infty$, each term in $I_N$ is roughly proportional to $1/N^2$, while the number of terms is proportional to $N$. Therefore, $I_N$ is roughly proportional to $1/N$ and goes to $0$ in the limit. This is illustrated in Figure \ref{figCBI}. 

Proceeding with the formal proof, omitting the superscript $|\phi\ra_{AB}$ for notational simplicity,
\begin{align}
|\pr_{\lam}(A_0 = 1) - \pr_{\lam}(A_0 = -1)| 	&= |\pr_{\lam}(A_{N,0} = 1) - \pr_{\lam}(A_{N,2N} = 1)|\notag\\
				&\le \sum_{|a-b|=1} |\pr_{\lam}(A_{N,a} = 1) - \pr_{\lam}(B_{N,b} = 1)|\notag\\
				&\le \sum_{|a-b|=1} \pr_{\lam}(A_{N,a} \ne B_{N,b}).\label{Sec2DerivBegin}
\end{align}
The first inequality is a simple triangle inequality, while the second follows from an inequality in probability theory:
\begin{align}
\forall z \in \mathrm{Range}(X) \cap \mathrm{Range}(Y)\!&:\notag\\
|\pr(X=z) - \pr(Y=z)| &= |\pr(X=z, Y=z) + \pr(X=z, Y \ne z)\notag\\
&- \pr(X=z, Y=z) - \pr(X \ne z, Y = z)|\notag\\
&\le \pr(X = z, Y \ne z) + \pr(X \ne z, Y = z) \notag\\
&\le \pr(X \ne Y).\label{ProbIneq}
\end{align}
Note that we have assumed ParInd here: this gaurantees that, for example, $\pr_{\lam}(A_{N,a} = 1)$ is well-defined and independent of what measurement is being performed on $B$.

Integrating both sides of \eqref{Sec2DerivBegin} using the measure $\mu(\lam)$ gives, using CompQuant,
\begin{align}
\int_\Lambda\mathrm{d}\mu(\lam)|\pr_{\lam}(A_0 = 1) - \pr_{\lam}(A_0 = -1)|	&\le \int_\Lambda\mathrm{d}\mu(\lam) \sum_{|a-b|=1} \pr_{\lam}(A_{N,a} \ne B_{N,b})\notag\\
							&= \sum_{|a-b|=1} \pr(A_{N,a} \ne B_{N,b}) = I_N.\label{IntIneq}
\end{align}
$I_N$ is a sum of quantum-mechanical probabilities and can therefore be calculated:
\begin{align}
a \in \mathcal{A}_N, b \in \mathcal{B}_N, |a - b| &= 1 \Rightarrow \pr^{|\phi\ra_{AB}} (A_{N,a} = \pm 1, B_{N,b} = \mp 1) = 1/2 \sin^2 \lt( \frac{\pi}{4N} \rt),\notag\\
\mbox{so }I_N &= 2N\sin^2 \lt( \frac{\pi}{4N} \rt) \le \frac{\pi^2}{8N} \quad \Rightarrow \quad \lim_{N\to\infty}I_N = 0.
\end{align}
Since \eqref{IntIneq} has to hold for all $N$, and the left-hand side is non-negative, it follows that
\begin{align}
\int_\Lambda\mathrm{d}\mu(\lam)|\pr_{\lam}(A_0 = 1) - \pr_{\lam}(A_0 = -1)| = 0  \quad \Rightarrow \quad \pr_{\lam}(A_0 = 1)\iae \pr_{\lam}(A_0 = -1).\label{Sec2DerivEnd}
\end{align}
Since both probabilities add up to $1$, they both equal $1/2$ almost everywhere. This can be written as\footnote{The observable need not be specified here because in a 2-dimensional Hilbert space a single projector fixes the other projector associated with the measurement.}
\begin{align}
\pr_{\lam}\lt([0]^A\rt) \iae \pr\lt([0]^A\rt) = 1/2;\notag\\
\pr_{\lam}\lt([1]^A\rt) \iae \pr\lt([1]^A\rt) = 1/2.\label{SecBellResPrj}
\end{align}
In fact, for any other basis $\{|0'\ra_A, |1'\ra_A\}$ of $\mathcal{H}_A$, there is a basis $\{|0'\ra_B, |1'\ra_B\}$ of $\mathcal{H}_B$ such that $|\phi\ra_{AB} = \frac1{\sqrt2}(|0'\ra_A|0'\ra_B + |1'\ra_A|1'\ra_B)$.\footnote{Namely, the basis $\{|0'\ra_B, |1'\ra_B\}$, where $|i'\ra_B := U^*_{ij}|j\ra_B$ and $U_{ij}$ is the unitary matrix such that $U_{ij}|j\ra_A = |i'\ra_A$.} Therefore, \eqref{SecBellResPrj} also holds for measurements on $A$ in any other basis. We have proved triviality for the state $|\phi\ra_{AB}$ \eqref{BellState}, in the 4-dimensional Hilbert space $\mathbb{C}^2 \otimes \mathbb{C}^2$.
%
%

\section{Generalizing to higher dimensional state spaces}\label{SecDDim}
\noindent The result of the previous section can easily be generalized to maximally entangled states in higher dimensional state spaces. Consider a bipartite system $AB$ in the state
\begin{align}
|\phi_d\ra_{AB} := \sum_{i=0}^{d-1} c_i |i\ra_A |i\ra_B \quad \in \mathcal{H}_A \otimes \mathcal{H}_B = \mathbb{C}^d \otimes \mathbb{C}^d,\label{MultiPhi}
\end{align}
with $c_i \in \mathbb{R}^+$ for all $i$, $d > 2$, and $\{|i\ra_A\}_{i=0}^{d-1}$ and $\{|i\ra_B\}_{i=0}^{d-1}$ orthonormal bases. Consider a measurement on $A$ in the basis $\{|i\ra_A\}_{i=0}^{d-1}$. We shall prove that equal coefficients imply equal $\lam$-probabilities:
\begin{align}
\forall j,k \in \mathbb{N}_d\!: \quad c_j = c_k \; \Rightarrow \; \pr_\lam\lt([j]^A\rt) \iae \pr_\lam\lt([k]^A\rt).\label{DDimResult}
\end{align}
In particular, if all coefficients in \eqref{MultiPhi} equal $1\big/\!\sqrt{d}$, all outcome probabilities equal $1/d$ almost everywhere:
\begin{align}
\forall i  \in \mathbb{N}_d\!: \quad \pr_\lam\lt([i]^A \rt) \iae \frac1d = \pr\lt([i]^A\rt).
\end{align}

The proof is essentially the same as in the previous section: we focus on the two terms with equal coefficients and apply the same steps. Let $c_j$ and $c_k$ be two equal coefficients. Similarly to the definitions \eqref{DefOtheta} and \eqref{AiBjDef} in the previous section, we define
\begin{align}
|\theta\ra &:= \cos(\theta/2)|j\ra + \sin(\theta/2)|k\ra, \quad \theta \in [0, \pi];\notag\\
\hat{O}_\theta &:= -1 \cdot [\theta] +1 \cdot [\theta + \pi] + \sum_{i \in \mathbb{N}_r \setminus \{j,k\}}(i + 2) \cdot [i]; \notag\\
\hat{A}_{N,a}	&:= \hat{O}^A_{a\pi/2N}, \quad a \in \mathcal{A}_N;\notag\\
\hat{B}_{N,b}	&:= \hat{O}^B_{b\pi/2N}, \quad b \in \mathcal{B}_N;\notag\\
\hat{A}_0	&:= \hat{A}_{N,0};\notag\\
I'_N &:= \sum_{|a-b|=1} \pr^{|\phi_d\ra_{AB}}(A_{N,a} \ne B_{N,b}).\label{DefsMulti}
\end{align}
Note that the observables $\hat{A}_{N,0}$ and $\hat{A}_{N,2N}$ correspond to the measurement in the basis $\{|i\ra_A\}_{i=0}^{d-1}$. Also, these observables are equal up to the exchange of the eigenvalues $+1$ and $-1$, which means that $A_{N,0} = \pm 1$ iff $A_{N,2N} = \mp 1$. The value of $I'_N$ is similar to that of $I_N$ \eqref{DefIN} in the previous section. The difference comes from the fact that the coefficients of the terms involving $|j\ra_A$ and $|k\ra_A$ equal $c_j$ instead of $1\big/\!\sqrt2$:
\begin{align}
I'_N = 4Nc_j^2 \sin^2 \lt( \frac{\pi}{4N} \rt) \le \frac{\pi^2c_j^2}{4N}.
\end{align}
Since $c_j$ is constant, $I'_N$, like $I_N$, has the property $\lim_{N \to \infty} I'_N = 0$. Repeating the steps \eqref{Sec2DerivBegin}--\eqref{Sec2DerivEnd} gives
\begin{align}
\pr_\lam\lt(A_0 = 1\rt) \iae \pr_\lam\lt(A_0 = -1\rt).
\end{align}
Using projectors, this can be expressed as
\begin{align}
\pr_\lam\lt([j]^A\rt)_{A_0} \iae \pr_\lam\lt([k]^A\rt)_{A_0}.\label{SomeFootnoteEq1}
\end{align}
Now, using the perfect correlation property \eqref{PC0} proven in Appendix \ref{AppPC}, setting $I=\{j\}$,
\begin{align}
\pr_\lam \lt( [j]^A \rt) \iae \pr_\lam \lt( [j]^B \rt),
\end{align}
we get
\begin{align}
\pr_\lam \lt( [j]^A \rt) \iae \pr_\lam \lt( [j]^B \rt) \iae \pr_\lam \lt( [k]^B \rt) \iae \pr_\lam \lt( [k]^A \rt).\label{SomeFootnoteEq2}
\end{align}
Note that the perfect correlation also implies that these probabilities are independent of the observables.
We now focus on the case with all coefficients equal to $1\big/\!\sqrt{d}$. Then, for any basis $\{|i'\ra_A\}_{i=0}^{d-1}$ there is a basis $\{|i'\ra_B\}_{i=0}^{d-1}$ such that $|\phi_d\ra_{AB} = \sum_{i=0}^{d-1} \lt( 1\big/\!\sqrt{d} \rt) |i'\ra_A |i'\ra_B$.\footnote{Namely, the basis $\{|i'\ra_B\}_{i=0}^{d-1}$, where $|i'\ra_B := U^*_{ij}|j\ra_B$ and $U_{ij}$ is the unitary matrix such that $U_{ij}|j\ra_A = |i'\ra_A$.} It follows that for measurements in any basis $\{|i'\ra_A\}_{i=0}^{d-1}$ or $\{|i'\ra_B\}_{i=0}^{d-1}$:
\begin{align}
\forall i \in \mathbb{N}_d\!: \pr_\lam\lt([i']^A \rt) \iae \frac1d &= \pr\lt([i']^A\rt), \notag\\
\pr_\lam\lt([i']^B \rt) \iae \frac1d &= \pr\lt([i']^B\rt).
\end{align}
In this case triviality also holds for measurements involving multidimensional projectors (or, equivalently, observables with degenerate eigenvalues). Consider a measurement on $A$ involving the projector $\sum_{i \in J}[i]^A$, where $J \subset \mathbb{N}_d$, and a simultaneous measurement on $B$ in the basis $\{|i\ra_B\}_{i=0}^{d-1}$. Then,
\begin{align}
\pr_\lam\lt( \sum_{i \in J} [i]^A \rt) \iae \sum_{i \in J} \pr_\lam\lt( [i]^B \rt) \iae \sum_{i \in J} \frac1d = \frac{\#J}d = \pr\lt( \sum_{i \in J} [i]^A \rt).\label{MultiProjResult}
\end{align}
Here we have made use of the perfect correlation property
\begin{align}
\pr_\lam^{|\phi_d\ra_{AB}} \lt( \sum_{i \in J} [i]^A \rt) \iae \sum_{i \in J} \pr_\lam^{|\phi_d\ra_{AB}}  \lt( [i]^B \rt),\label{PC0}
\end{align}
also proven in Appendix \ref{AppPC}. Remember that by ParInd, the outcome probabilities of a measurement performed on $A$ are independent of whether any measurement is performed on $B$. Therefore, while we did consider a specific measurement on $B$ in \eqref{MultiProjResult}, the result
\begin{align}
\pr_\lam\lt( \sum_{i \in J} [i]^A \rt) \iae \pr\lt( \sum_{i \in J} [i]^A \rt)
\end{align}
holds in general.

\section{Generalizing to coefficients $c_i^2 \in \mathbb{Q}$}\label{SecSqrRat}
\noindent In this section the result of the previous sections is generalized to states of which the Schmidt coefficients equal square roots of rational numbers:\footnote{The set of rational numbers is denoted by $\mathbb{Q}$.}
\begin{align}
|\phi_d\ra_{AB} := \sum_{i=0}^{d-1} c_i |i\ra_A |i\ra_B \quad \in \mathcal{H}_A \otimes \mathcal{H}_B,\label{RatPhi}
\end{align}
where $c_i^2 \in \mathbb{Q}$ and $c_i > 0$  for all $i$, $d > 2$, $\mathcal{H}_A$ and $\mathcal{H}_B$ are both at least of dimension $d$, and $\{|i\ra_A\}_{i=0}^{d-1}$ and $\{|i\ra_B\}_{i=0}^{d-1}$ are sets of orthonormal vectors.\footnote{Since we allow for state spaces of dimension larger than $d$, $\{|i\ra_A\}_{i=0}^{d-1}$ and $\{|i\ra_B\}_{i=0}^{d-1}$ do not necessarily span $\mathcal{H}_A$ and $\mathcal{H}_B$ and therefore might not be bases.} We shall prove that also in this case the $\lam$-probabilities equal the quantum probabilities:
\begin{align}
\forall i  \in \mathbb{N}_d\!: \quad \pr^{|\phi_d\ra_{AB}}_\lam\lt([i]^A \rt) \iae \pr^{|\phi_d\ra_{AB}}\lt([i]^A\rt).
\end{align}

The proof in this section is based on the following idea. System $AB$ can be coupled to another system $A'B'$ such that the combined system $AA'BB'$ is approximately in a maximally entangled state. More precisely, for each $i$ the term $|i\ra_A|i\ra_B$ is coupled to an approximate entangled state with a number of terms that is proportional to $c_i^2$. To find these numbers, we consider the common denominator $r$ of all fractions $\{c_i^2\}_{i=0}^{d-1}$, such that we can write $c_i^2 = m_i/r$ with $m_i \in \mathbb{N}$. The numerators $m_i$ are then proportional to $c_i^2$. Then, for measurements on $AA'BB'$, which is approximately in a maximally entangled state, the result of the previous section can be applied. It turns out that this also puts constraints on the $\lam$-probabilities for measurements on system $AB$ alone. 

To produce the maximally entangled state, we assume the presence of yet another bipartite system $A''B''$ that is prepared in a special `embezzling state'. The family of embezzling states was first introduced in \cite{Dam2003}, and they have the special property that any bipartite state can be approximately extracted from it using local unitary operations (i.e. by applying separate unitary transformations on $A''$ and $B''$). The precision of this operation can be enhanced by using an embezzling state of higher dimension.

Because we have to use the notion of `approximate states' in this section, we define a distance measure for quantum states, called the trace distance:
\begin{align}
D(|\psi\ra,|\phi\ra) &:= (1/2)\mathrm{Tr}\Bigl| [\psi] - [\phi] \Bigr|,\mbox{ where}\label{TraceProb}\\
|A| &:= \sqrt{A^\dagger A}.\notag
\end{align}
As shown in \cite{Niel2000}, the trace distance is a metric, and it provides an upper bound for the difference in quantum probabilities for the outcome associated with any projector $\hat{P}$:
\begin{align}
\lt| \la\Psi|\hat{P}|\Psi\ra - \la\phi|\hat{P}|\phi\ra \rt| \le D(|\psi\ra,|\phi\ra).\label{StateProbTD}
\end{align}
Also, for pure states the following relation holds between the trace distance and another distance measure, the fidelity $\mathcal{F}$:
\begin{align}
D(|\psi\ra,|\phi\ra) &= \sqrt{1 - \mathcal{F}(|\psi\ra,|\phi\ra)^2}, \notag\\
\mbox{where } \mathcal{F}(|\psi\ra,|\phi\ra) &:= |\la \psi | \phi \ra|.\label{RelFidTD}
\end{align}
Beginning with the proof, first define\footnote{``LCD'' stands for ``least common denominator''}
\begin{align}
r		&:= \mathrm{LCD}\lt(\lt\{c_i^2\rt\}_{i=0}^{d-1} \rt);\notag\\
m_i		&:= r c^2_i.\label{dmrc}
\end{align}
The family of embezzling states $\{ |\tau_n\ra \}_{n=1}^\infty$ is defined as
\begin{align}
|\tau_n\ra := \frac1{\sqrt{C_n}}\sum_{j=0}^{n-1}\frac1{\sqrt{j+1}}|j\ra \otimes |j\ra \quad \in \mathbb{C}^n \otimes \mathbb{C}^n.\label{EmbState}
\end{align}
$C_n := \sum_{j=0}^{n-1} 1/{(j+1)}$ is a normalization constant, and $\{|j\ra\}_{j=0}^{n-1}$ an orthonormal basis of $\mathbb{C}^n$. System $A''B''$ is prepared in one of these states, indexed by $n$. Another system $A'B'$ is prepared in a simple product state
\begin{align}
|0\ra_{A'}|0\ra_{B'} \in \mathcal{H}_{A'} \otimes \mathcal{H}_{B'},
\end{align}
where $\mathcal{H}_{A'}$ and $\mathcal{H}_{B'}$ are at least of dimension $\mathrm{max}_{i \in \mathbb{N}_d} m_i$. Taking all systems together, the state before the measurement is
\begin{align}
|\Psi_n\ra &:= |\tau_n\ra_{A''B''} \otimes |0\ra_{A'}|0\ra_{B'} \otimes |\phi_d\ra_{AB}.
\end{align}

Now, for each $i$, a maximally entangled state with Schmidt number $m_i$ is extracted from the embezzling state, and coupled to the term $|i\ra_A|i\ra_B$. In Appendix \ref{AppEmb} it is shown that there are unitary operators $U^{AA'A''}_n, U^{BB'B''}_n$ which perform this task with a precision that increases as $n \to \infty$:
\begin{align}
\lim_{n\to\infty} &D\lt(U^{AA'A''}_n \otimes U^{BB'B''}_n|\Psi_n\ra, |\chi_n\ra\rt) = 0,\mbox{ where}\label{Eq553}\\
|\chi_n\ra	&:= |\tau_n\ra_{A''B''} \otimes \sum_{i=0}^{d-1} \lt( \sqrt{\frac{m_i}r}|i\ra_A|i\ra_B\sum_{j=0}^{m_i-1} \frac1{\sqrt{m_i}}|j\ra_{A'}|j\ra_{B'}\rt)\notag\\
		&= |\tau_n\ra_{A''B''} \otimes \sum_{i=0}^{d-1} \sum_{j=0}^{m_i-1} \lt( \frac1{\sqrt{r}}|i\ra_A|i\ra_B |j\ra_{A'}|j\ra_{B'}\rt).\notag
\end{align}
Here, $\{|j\ra_{A'}\}_{j=0}^{m-1}$ and $\{|j\ra_{B'}\}_{j=0}^{m-1}$ are sets of orthonormal vectors.
Now consider a measurement that consists of first performing the above unitary transformation, resulting approximately in a maximally entangled state for the system $AA'BB'$, and then measuring this system in the basis $\{|i,j\ra_{AA'}\}$, where $|i,j\ra := |i\ra|j\ra$. Note that, if the state $U^{AA'A''}_n \otimes U^{BB'B''}_n|\Psi_n\ra$ is close to a maximally entangled state, by \eqref{StateProbTD} the outcome probabilities are also close to those of a maximally entangled state. Therefore, the result of the previous section can be applied. Actually, the unitary transformation can be included in the definition of the measurement projectors, because a unitary transformation applied to a complete set of orthogonal projectors gives another complete set of orthogonal projectors.

Define an index set containing all values of the pair $(i,j)$ occurring in $|\chi_n\ra$:
\begin{align}
I^{\mathrm{ind}} :=	\;&\Big\{(0,0), (0,1), \dots, (0,m_0-1),\notag\\
	&(1,0),\dots,(1,m_1 -1),\notag\\
	&\dots,\notag\\
	&(d-1,0), \dots, (d-1,m_{d-1} -1)\Big\}.
\end{align}
For $(i_1, j_1), (i_2, j_2) \in I^{\mathrm{ind}}$ define, similarly to \eqref{DefsMulti} in the previous section,\footnote{While the symbols defined in \eqref{manydefs} and \eqref{DefINn} depend on the pairs $(i_1, j_1)$ and $(i_2, j_2)$, these are not included as indices for notational simplicity.}
\begin{align}
|\theta\ra := &\cos ({\theta}/2)|i_1,j_1\ra + \sin({\theta}/2)|i_2, j_2\ra;\notag\\
\hat{O}_\theta := &-1\cdot[\theta] +1\cdot[\theta + \pi] + \hspace{-2.5em} \sum_{(i,j) \in I^{\mathrm{ind}}\setminus\{(i_1,j_1),(i_2, j_2)\}} \hspace{-2.5em} \lt( 2^i 3^j + 2 \rt) \cdot [i,j];\notag\\ %
\hat{A}_{N,n,a} &:= \lt(U^{AA'A''}_n\rt)^{-1} \lt( {\mathbb{I}}^{A''} \otimes \hat{O}^{AA'}_{a\pi/2N}\rt)\lt(U^{AA'A''}_n\rt),	\quad a \in \mathcal{A}_N;\notag\\
\hat{B}_{N,n,b} &:= \lt(U^{BB'B''}_n\rt)^{-1} \lt( {\mathbb{I}}^{B''} \otimes \hat{O}^{BB'}_{b\pi/2N}\rt)\lt(U^{BB'B''}_n\rt),	\quad b \in \mathcal{B}_N;\notag\\
\hat{A}_{n,0} &:=\hat{A}_{N,n,0}.\label{manydefs}
\end{align}
Note that, as in the previous section, $A_{n,0} = \pm 1$ iff $A_{N,n,2N} = \mp 1$. The use of the expression $(2^i3^j + 2)$ in the definition of $\hat{O}_\theta$ guarantees that distinct eigenvalues are assigned to every projector $[i,j]$, as in the previous section. Note that the observables $\hat{A}$ and $\hat{B}$ include the unitary operators $U^{AA'A''}_n$ and $U^{BB'B''}_n$, and therefore depend on $n$, which corresponds to the precision of the embezzlement transformation.

Finally, a correlation measure is again defined, which now also depends on the precision $n$:
\begin{align}
I_{N,n} &:= \sum_{|a-b|=1} \pr^{|\Psi_n\ra} \lt(A_{N,n,a} \ne B_{N,n,b}\rt).\label{DefINn}
\end{align}
This quantity has the property, similar to $I_N$ in the previous section, that it vanishes when taking the limits $n \to \infty$ and then $N \to \infty$:
\begin{align}
\forall (i_1, j_1), (i_2, j_2) \in I^{\mathrm{ind}} \!: \lim_{N \to \infty} \lim_{n \to \infty} I_{N,n} = 0.\label{INnLim}
\end{align}
The proof of \eqref{INnLim} can be found in Appendix \ref{SubINn}. The correctness is already suggested by the fact that $I_{N,n}$ consists of $2N$ probabilities, and as mentioned above, those probabilities get closer to probabilities for a maximally entangled state as $n$ increases.

Analogously to the previous two sections, we have
\begin{align}
\lt|\pr^{|\Psi_n\ra}_\lam\lt(A_{n,0} = 1\rt) - \pr^{|\Psi_n\ra}_\lam\lt(A_{n,0} = -1\rt)\rt| 	&= \lt|\pr^{|\Psi_n\ra}_\lam\lt(A_{N,n,0} = 1\rt) - \pr^{|\Psi_n\ra}_\lam\lt(A_{N,n,2N} = 1\rt)\rt|\notag\\
						&\le \sum_{|a-b|=1} \lt|\pr^{|\Psi_n\ra}_\lam\lt(A_{N,n,a} = 1\rt) - \pr^{|\Psi_n\ra}_\lam\lt(B_{N,n,b} = 1\rt)\rt|\notag\\
						&\le \sum_{|a-b|=1} \pr^{|\Psi_n\ra}_\lam\lt(A_{N,n,a} \ne B_{N,n,b}\rt).
\end{align}
Integrating with the measure $\mu(\lam)$ gives
\begin{align}
\int_\Lambda\mathrm{d}\mu(\lam) \lt|\pr^{|\Psi_n\ra}_\lam\lt(A_{n,0} = 1\rt) - \pr^{|\Psi_n\ra}_\lam\lt(A_{n,0} = -1\rt)\rt| \le I_{N,n}.
\end{align}
Let $\epsilon > 0$. By \eqref{INnLim}, we can choose $n, N$ such that for all $(i_1, j_1), (i_2, j_2) \in I^{\mathrm{ind}}$, $I_{N,n} < \epsilon$, so that
\begin{align}
\forall (i_1, j_1), (i_2, j_2) \in I^{\mathrm{ind}}\!: \int_\Lambda\mathrm{d}\mu(\lam) \lt|\pr^{|\Psi_n\ra}_\lam\lt(A_{n,0} = 1 \rt) - \pr^{|\Psi_n\ra}_\lam\lt(A_{n,0} = -1\rt)\rt| < \epsilon.\label{SecResult}
\end{align}
Defining the projector
\begin{align}
\hat{E}_{(i,j),n} &:= \lt(U^{AA'A''}_n\rt)^{-1}\lt({\mathbb{I}}^{A''} \otimes [i,j]^{AA'}\rt) \lt(U^{AA'A''}_n\rt),
\end{align}
and noting that
\begin{align}
\hat{A}_{n,0} = -1 \cdot \hat{E}_{(i_1,j_1),n} +1 \cdot \hat{E}_{(i_2,j_2),n} + \dots,
\end{align}
we get, switching to the notation with projectors,
\begin{align}
\forall (i_1, j_1), (i_2, j_2) \in I^{\mathrm{ind}}\!: \int_\Lambda\mathrm{d}\mu(\lam) \lt|\pr^{|\Psi_n\ra}_\lam\lt(\hat{E}_{(i_2, j_2),n}\rt)_{\hat{A}_{n,0}} - \pr^{|\Psi_n\ra}_\lam\lt(\hat{E}_{(i_1,j_1),n}\rt)_{\hat{A}_{n,0}} \rt| < \epsilon.\label{Eqij1ij2}
\end{align}
%
%
By CompQuant, we have
\begin{align}
\sum_{(i',j') \in I^{\mathrm{ind}}} \int_\Lambda \mathrm{d}\mu(\lam) \pr^{|\Psi_n\ra}_\lam\lt(\hat{E}_{(i',j'),n} \rt) = 1 \quad \Rightarrow \quad \sum_{(i',j') \in I^{\mathrm{ind}}}\pr^{|\Psi_n\ra}_\lam\lt(\hat{E}_{(i',j'),n} \rt) \iae 1.
\end{align}
Using this, and a triangle inequality, we get
\begin{align}
&\forall (i,j) \in I^\mathrm{ind}\!: \int_\Lambda\mathrm{d}\mu(\lam) \lt| \pr^{|\Psi_n\ra}_\lam\lt(\hat{E}_{(i,j),n} \rt)_{\hat{A}_{n,0}} - \frac1r \rt|\notag\\
&= \frac1r \int_\Lambda\mathrm{d}\mu(\lam) \lt| r \cdot \pr^{|\Psi_n\ra}_\lam\lt(\hat{E}_{(i,j),n} \rt)_{\hat{A}_{n,0}} - \sum_{(i',j') \in I^{\mathrm{ind}}}\pr^{|\Psi_n\ra}_\lam\lt(\hat{E}_{(i',j'),n} \rt)_{\hat{A}_{n,0}}\rt|\notag\\
&\le \frac1r \sum_{(i',j') \in I^{\mathrm{ind}}}\int_\Lambda\mathrm{d}\mu(\lam) \lt|\pr^{|\Psi_n\ra}_\lam\lt(\hat{E}_{(i,j),n} \rt)_{\hat{A}_{n,0}} - \pr^{|\Psi_n\ra}_\lam\lt(\hat{E}_{(i',j'),n} \rt)_{\hat{A}_{n,0}} \rt| < \frac{\#I^{\mathrm{ind}}}{r} \epsilon = \epsilon.
\end{align}
Using a triangle inequality once more,
\begin{align}
\forall i \in\mathbb{N}_d \!: \int_\Lambda\mathrm{d}\mu(\lam) \lt|\lt(\sum_{j=0}^{m_i-1} \pr^{|\Psi_n\ra}_\lam\lt(\hat{E}_{(i,j),n} \rt)\rt)_{\hat{A}_{n,0}} - \frac{m_i}r \rt|\notag\\
\le \sum_{j=0}^{m_i-1} \int_\Lambda\mathrm{d}\mu(\lam) \lt| \pr^{|\Psi_n\ra}_\lam\lt(\hat{E}_{(i,j),n} \rt)_{\hat{A}_{n,0}} - \frac1r \rt| < m_i\epsilon.
\end{align}
Using the correlation property
\begin{align}
\sum_{j=0}^{m_{i}-1}\pr^{|\Psi_n\ra}_\lam\lt(\hat{E}_{(i,j),n}\rt) &\iae \pr^{|\Psi_n\ra}_\lam\lt([i]^B\rt),\label{CorProp}
\end{align}
proven in Appendix \ref{AppPC}, we arrive at
\begin{align}
\forall i \in\mathbb{N}_d \!: \int_\Lambda\mathrm{d}\mu(\lam) \lt| \lt( \sum_{j=0}^{m_i-1} \pr^{|\Psi_n\ra}_\lam\lt(\hat{E}_{(i,j),n} \rt) \rt)_{\hat{A}_{n,0}} - \frac{m_i}r \rt| \notag\\
= \int_\Lambda\mathrm{d}\mu(\lam) \lt|\pr^{|\Psi_n\ra}_\lam\lt([i]^B \rt) - \frac{m_i}r \rt| < m_i\epsilon.\label{PrevEq}
\end{align}
Since \eqref{PrevEq} holds for any $\epsilon > 0$, we get, also applying \eqref{PE} by noting that $|\Psi_n\ra = |\phi_d\ra_{AB} \otimes \ldots$,
\begin{align}
\int_\Lambda\mathrm{d}\mu(\lam) \lt|\pr^{|\phi_d\ra_{AB}}_\lam\lt([i]^B \rt) - \frac{m_i}r \rt| &= 0 \notag\\
\Rightarrow \quad \pr^{|\phi_d\ra_{AB}}_\lam\lt([i]^B \rt) \iae\frac{m_i}r = c_i^2 &= \pr^{|\phi_d\ra_{AB}}\lt([i]^B \rt),\label{Sec4Result}
\end{align}
which is the desired result.

Note that because of ParInd, once we have derived results for the probabilities for measurements on system $B$ using the correlation property \label{CorProp} and certain measurements performed on system $AA'A''$, these results still hold when other measurements, or no measurements at all, are being performed on system $AA'A''$.

Also note that, starting with the system $AA'A''BB'B''$, with $A''B''$ prepared in an embezzling state, we have reached a conclusion about measurements on $AB$, prepared in the pure state \eqref{RatPhi}. Above we have applied \eqref{PE}, which states that the probabilities $\pr^{|\phi_d\ra_{AB}}_\lam\lt([i]^B\rt)$ are independent of the states of the systems $A''B''$ and $A'B'$. Therefore, while these systems were used for the derivation of \eqref{Sec4Result}, it follows that the result also holds without considering the systems $A''B''$ and $A'B'$ in specially prepared states.

Like in the previous section, the result also holds with $A$ and $B$ interchanged, and also for degenerate measurements.

\section{Generalizing to arbitrary coefficients}\label{SecAny}
\noindent Now suppose we have a state of the form
\begin{align}
|\phi_d\ra_{AB} := \sum_{i=0}^{d-1} c_i |i\ra_A |i\ra_B \quad \in \mathcal{H}_A \otimes \mathcal{H}_B,
\end{align}
where $c_i \in \mathbb{R}^+$ for all $i$, $d > 2$, $\mathcal{H}_A$ and $\mathcal{H}_B$ are both at least of dimension $d$, and $\{|i\ra_A\}_{i=0}^{d-1}$ and $\{|i\ra_B\}_{i=0}^{d-1}$ are sets of orthonormal vectors. Note that any state can be written in this form by Schmidt's decomposition theorem. Since we cannot generally write $c_i^2 = m_i/r$ with $m_i, r \in \mathbb{N}$, the strategy of the previous section can not be applied. However, because the set of rational numbers is dense in the set of real numbers, the numbers $c_i^2$ can be approximated with rational numbers $c_{i,l}^2$ by defining the sequences
\begin{align}
(c_{i,l})_{l=1}^\infty, \mbox{ with }	&\forall i ,l \in \mathbb{N} \!: c_{i,l}^2 \in \mathbb{Q},\\
				&\forall i \!: \lim_{l \to \infty} c_{i,l} = c_i,\label{Limcil}\\
				&\forall l \!: \sum_{i=0}^{d-1} c_{i,l}^2 = 1.\label{CilNorm}
\end{align}
As in the previous section, the idea is to apply a unitary transformation that results approximately in a maximally entangled state. But since the $c_i^2$ are not rational numbers, we have to use the rational numbers $c_{i,l}^2$ to determine the numerators and denominators for the unitary transformations. This time, to achieve any desired precision, not only $n$ must be chosen large enough, but also $l$. Again, the unitary transformation will be included in the measurement operators.

Instead of focussing on two terms, like we did in the previous section, the terms of the approximately maximally entangled state are now partitioned into two sets of equal size. Then, a measurement is considered using a projector that projects on the space spanned by one such set of terms, and measurements where this projector is `rotated' by an angle $\theta$. For a maximally entangled state, the quantum-mechanical statistics for such measurements are equal to those of the measurements on Bell states, considered in Section \ref{SecBell}. Since we have a state that is close to a maximally entangled state, by \eqref{StateProbTD} the quantum probabilities are also close to those of Bell states, i.e. close to $1/2$. Using a similar derivation as in the previous two sections, it can be shown that also the $\lam$-probabilities are close to $1/2$. Then, using a lemma proved in Appendix \ref{AppLemma}, we can also derive $\lam$-probabilities for projectors $[i]^A$. In fact, those probabilities are again equal to the quantum-mechanical ones for almost every $\lam \in \Lambda$.

For each $l \in \mathbb{N}$, define $r_l$ as the least \emph{even} common denominator of the fractions $\{c^2_{i,l}\}_{i=0}^{d-1}$,\footnote{This is achieved by including $1/2$ in the set of fractions.} and let $\{m_{i,l}\}_{i=0}^{d-1}$ be the corresponding numerators:
\begin{align}
r_l &:= \mathrm{LCD}\lt(\lt\{c_{i,l}^2\rt\}_{i=0}^{d-1} \cup \{1/2\}\rt);\notag\\
m_{i,l} &:= r_l c^2_{i,l}.\label{Idmrc}
\end{align}
The reason that $r_l$ needs to be even is that an approximately maximally entangled state with an even number of terms is needed in order to be able to divide the terms into two groups of equal size. 

As in the previous section, we assume the presence of two extra bipartite systems, of which one is prepared in an embezzling state \eqref{EmbState}:
\begin{align}
|\Psi_n\ra &:= |\tau_n\ra_{A''B''} \otimes |0\ra_{A'}|0\ra_{B'} \otimes |\phi_d\ra_{AB}.
\end{align}
Note that, for given $l$, $\mathcal{H}_{A'}$ and $\mathcal{H}_{B'}$ are both at least of dimension $\mathrm{max}_{i \in \mathbb{N}_d} m_{i,l}$.

Again, define an index set:
\begin{align}
J^{\mathrm{ind}}_l :=	 \: &\Big\{(0,0), (0,1), \dots, (0,m_{0,l}-1),\notag\\
	&(1,0),\dots,(1,m_{1,l} -1),\notag\\
	&\dots,\notag\\
	&(d-1,0), \dots, (d-1,m_{d-1,l} -1)\Big\}.
\end{align}
Note that $\sum_{i=0}^{d-1} m_{i,l} = \sum_{i=0}^{d-1} r_l c_{i,l}^2 = r_l$ by \eqref{Idmrc} and \eqref{CilNorm}, and therefore $\#J^{\mathrm{ind}}_l = r_l$.

Let $J_l \subset J_l^{\mathrm{ind}}$ with $\#J_l = r_l/2$; and $p_l: J_l \rightarrow J^{\mathrm{ind}}_l\!\setminus\! J_l$ an arbitrary bijection. Then, define
\begin{align}
\hat{O}^{AA'}_{J_l,\theta} &:= +1 \cdot \sum_{(i,j) \in J_l} \lt[\cos (\theta/2)|i,j\ra_{AA'} + \sin(\theta/2)|p_l(i,j)\ra_{AA'}\rt]\notag\\
&\quad\quad - 1 \cdot \lt({\mathbb{I}}^{AA'} - \sum_{(i,j) \in J_l} \lt[\cos (\theta/2)|i,j\ra_{AA'} + \sin(\theta/2)|p_l(i,j)\ra_{AA'}\rt]\rt);\notag\\
\hat{A}_{J_l, N, n, l, a} &:= \lt(U^{AA'A''}_{n,l}\rt)^{-1} \lt({\mathbb{I}}^{A''} \otimes \hat{O}^{AA;}_{J_l, a\pi/2N}\rt) \lt(U^{AA'A''}_{n,l}\rt), \quad a \in \mathcal{A}_N\!\setminus\!\{2N\},
\end{align}
and $\hat{O}^{BB'}_{J_l,\theta}, \hat{B}_{J_l, N, n, l, b}$ analogously. Furthermore, define
\begin{align}
\hat{A}_{J_l,n,l,0} &:= \hat{A}_{J_l,N,n,l,0};\notag\\
\hat{A}_{J_l,N,n,l,2N} &:= -\hat{A}_{J_l,n,l,0}.
\end{align}
Now the correlation measure can be defined:
\begin{align}
I_{J_l, N, n, l} := \sum_{|a-b|=1} \pr^{|\Psi_n\ra}\lt(A_{J_l, N, n, l, a} \ne B_{J_l, N, n, l, b}\rt).
\end{align}
As shown in Appendix \ref{SubIJL}, this measure has the property
\begin{align}
\lim_{N \to \infty} \lim_{l \to \infty} \lim_{n \to \infty} I_{J_l, N, n, l} = 0.\label{LimIJNnl}
\end{align}
Now,
\begin{align}
2 \lt| \pr^{|\Psi_n\ra}_\lam\lt(A_{J_l, n, l, 0} = +1\rt) - 1/2 \rt| 	&= \lt| \pr^{|\Psi_n\ra}_\lam\lt(A_{J_l, n, l, 0} = +1\rt) - \pr^{|\Psi_n\ra}_\lam\lt(A_{J_l,n,l,0} = -1\rt) \rt| \notag\\
&= \lt| \pr^{|\Psi_n\ra}_\lam\lt(A_{J_l, N, n, l, 0} = +1\rt) - \pr^{|\Psi_n\ra}_\lam\lt(A_{J_l, N, n, l, 2N} = +1\rt) \rt| \notag\\
&\le \sum_{|a-b|=1} \lt| \pr^{|\Psi_n\ra}_\lam\lt(A_{J_l, N, n, l, a} = +1 \rt) - \pr^{|\Psi_n\ra}_\lam\lt(B_{J_l, N, n, l, b} = +1 \rt) \rt| \notag\\
&\le \sum_{|a-b|=1} \pr^{|\Psi_n\ra}_\lam\lt(A_{J_l, N, n, l, a} \ne B_{J_l, N, n, l, b}\rt)
\end{align}
Integrating with the measure $\mu(\lam)$ gives
\begin{align}
\int_\Lambda\mathrm{d}\mu(\lam) \lt| \pr^{|\Psi_n\ra}_\lam\lt(A_{J_l, n, l, 0} = +1\rt) -1/2 \rt| \le (1/2) I_{J_l, N, n, l}.
\end{align}
Let $\epsilon > 0$. By \eqref{LimIJNnl} and \eqref{Limcil} we can choose $N, l, n \in \mathbb{N}$ such that for all $J_l \subset J_l^{\mathrm{ind}}$ with $\#J_l = r_l/2$,
\begin{align}
\int_\Lambda\mathrm{d}\mu(\lam) \lt| \pr^{|\Psi_n\ra}_\lam\lt(A_{J_l, n, l, 0} = +1\rt) -1/2 \rt| &< \epsilon;\notag\\
\forall i \in \mathbb{N}_d \!: |c_{i,l}^2 - c_i^2| &< \epsilon.
\end{align}
Now, define the projectors
\begin{align}
\hat{E}_{(i,j),n,l} &:= \lt(U^{AA'A''}_{n,l}\rt)^{-1} \lt( [i,j]^{AA'} \otimes {\mathbb{I}}^{A''} \rt) \lt(U^{AA'A''}_{n,l} \rt),\notag\\
\hat{F}_{(i,j),n,l} &:= \lt(U^{BB'B''}_{n,l}\rt)^{-1} \lt( [i,j]^{BB'} \otimes {\mathbb{I}}^{B''} \rt) \lt(U^{BB'B''}_{n,l} \rt).
\end{align}
Noting that
\begin{align}
\hat{A}_{J_l, n, l, 0} = +1 \cdot \lt( \sum_{(i,j) \in J_l} \hat{E}_{(i,j),n,l} \rt) -1 \cdot \lt( \mathbb{I}^{AA'A''} - \sum_{(i,j) \in J_l} \hat{E}_{(i,j),n,l} \rt),
\end{align}
we switch to the notation with projectors, and using the correlation property \eqref{PC3},
\begin{align}
\pr^{|\Psi_n\ra}_\lam \lt(\sum_{(i,j) \in J_l} \hat{E}_{(i,j),n,l}\rt) &\iae \sum_{(i,j) \in J_l} \pr^{|\Psi_n\ra}_\lam \lt(\hat{F}_{(i,j),n,l}\rt),
\end{align}
we arrive at, for all $J_l \subset J_l^{\mathrm{ind}}$ with $\#J_l = r_l/2$, 
\begin{align}
\int_\Lambda\mathrm{d}\mu(\lam) \lt|\pr^{|\Psi_n\ra}_\lam\lt(\sum_{(i,j) \in J_l} \hat{E}_{(i,j), n, l} \rt) -1/2 \rt| &< \epsilon \notag\\
\Rightarrow \quad \int_\Lambda\mathrm{d}\mu(\lam) \lt| \sum_{(i,j) \in J_l} \pr^{|\Psi_n\ra}_\lam\lt(\hat{F}_{(i,j), n, l} \rt) -1/2 \rt| &< \epsilon.\label{Lem2AnteInsert}
\end{align}
Now, we need the following lemma, proven in Appendix \ref{AppLemma}:
\begin{lemma}
Let $\Lambda$ be a measurable space with measure $\mu$, let $r$ be an even, positive integer, and let
\begin{align}
\lt\{\lt(p^\lam_i\rt)_{i=0}^{r-1} \me| \lam \in \Lambda \rt\}
\end{align}
be some collection of sequences satisfying, for all $I \subset I^{\mathrm{ind}} := \mathbb{N}_r$ with $\#I = r/2$,
\begin{align}
\int_\Lambda\mathrm{d}\mu(\lam) \lt|\lt(\sum_{i \in I} p^\lam_i \rt) -1/2 \rt| &< \epsilon.\label{Lem2Ante}
\end{align}
Then for all $J \subset I^{\mathrm{ind}}$
\begin{align}
\int_\Lambda\mathrm{d}\mu(\lam) \lt| \lt(\sum_{i \in J} p^\lam_i \rt) - {\#J}/r \rt| &< 2\epsilon.
\end{align}
\end{lemma}
\noindent Applying this lemma to the right-hand side of \eqref{Lem2AnteInsert} with
\begin{align}
\lt(p_i^\lam\rt)_{i=0}^{r-1} \equiv \lt(\pr^{|\Psi_n\ra}_\lam\lt(\hat{F}_{(0,0), n, l} \rt), \pr^{|\Psi_n\ra}_\lam\lt(\hat{F}_{(0,1), n, l} \rt), \dots, \pr^{|\Psi_n\ra}_\lam\lt(\hat{F}_{(d-1,m_{d-1,l}), n, l} \rt)\rt),
\end{align}
we get, for all $i \in \mathbb{N}_d$,
\begin{align}
\int_\Lambda\mathrm{d}\mu(\lam) \lt| \sum_{j=0}^{m_{i,l} -1} \pr^{|\Psi_n\ra}_\lam\lt( \hat{F}_{(i,j), n, l} \rt) - \frac{m_{i,l}}{r_l} \rt| < 2\epsilon.
\end{align}
Again, applying perfect correlation \eqref{PC2},
\begin{align}
\sum_{j=0}^{m_{i,l} -1} \pr^{|\Psi_n\ra}_\lam\lt(\hat{F}_{(i,j),n,l} \rt) \iae \pr^{|\Psi_n\ra}_\lam\lt([i]^A\rt), 
\end{align}
and noting that $m_{i,l}/r_l = c_{i,l}^2$ by \eqref{Idmrc}, we obtain, also using \eqref{PE},
\begin{align}
\int_\Lambda\mathrm{d}\mu(\lam) \lt|\pr^{|\phi_d\ra_{AB}}_\lam\lt([i]^A\rt) - c^2_{i,l}\rt| < 2\epsilon. %
\end{align}
Finally, using the triangle inequality $|x - y| \le |x - z| + |z - y|$
\begin{align}
&\int_\Lambda\mathrm{d}\mu(\lam) \lt|\pr^{|\phi_d\ra_{AB}}_\lam\lt([i]^A\rt) - c^2_i \rt| \notag\\
&\le  \int_\Lambda\mathrm{d}\mu(\lam) \lt|\pr^{|\phi_d\ra_{AB}}_\lam\lt([i]^A\rt) - c^2_{i,l}\rt| +  \int_\Lambda\mathrm{d}\mu(\lam) \lt|c^2_{i,l} - c^2_i\rt| \notag\\
&< 3\epsilon,
\end{align}
and since the inequality holds for all $\epsilon > 0$, we get
\begin{align}
\int_\Lambda\mathrm{d}\mu(\lam) \lt|\pr^{|\phi_d\ra_{AB}}_\lam\lt([i]^A\rt) - c^2_i\rt| = 0\notag\\
\Rightarrow \pr^{|\phi_d\ra_{AB}}_\lam\lt([i]^A\rt) \iae c_i^2 = \pr^{|\phi_d\ra_{AB}}\lt([i]^A\rt),
\end{align}
which is the desired result. Again, the result also holds with $A$ and $B$ interchanged, and also for degenerate measurements.

\section{Generalizing to any measurement}\label{SecAnyProj}
In the above sections, we considered measurements on subsystems of bipartite systems in entangled states. However, since in fact every measurement involves such entanglement, something can also be said about general measurements. In this section, we show how this can be done.\footnote{The results of this section are largely based on a suggestion made by Guido Bacciagaluppi (private communication).} This generalization is performed step-by-step, by considering three types of measurements in turn: measurements of the first kind, measurements of the second kind, and POVM\footnote{POVM stands for Positive Operator-Valued Measure. A detailed treatment of these different types of measurements can be found in \citet{Busc1996}.} measurements.
\subsection{Measurements of the first kind}
The measurement process can be divided into two steps. First, the system to be measured is coupled to the measurement apparatus, which results in entanglement between the system and the apparatus. Then, the measurement result is read off the measurement apparatus. In the case of a measurement of the first kind with projectors $\{\hat{E}^A_j\}_{j=0}^{d-1}$, the coupling has the following form:
\begin{align}
|\psi\ra_A|0\ra_B = \sum_{j=0}^{d-1} \hat{E}^A_j |\psi\ra_A |0\ra_B \mapsto \sum_{j=0}^{d-1} \hat{E}^A_j |\psi\ra_A |j\ra_B = \sum_{j=0}^{d-1} c_j |j\ra_A |j\ra_B,
\end{align}
where
\begin{align}
|j\ra_A := \frac{\hat{E}^A_j|\psi\ra_A}{\lt\|\hat{E}^A_j|\psi\ra_A\rt\|} \quad \mbox{, and } \quad c_j := |\la j|\psi\ra_A| = \lt\|\hat{E}^A_j|\psi\ra_A\rt\|,
\end{align}
$|i\ra_B$ are (mutually orthogonal) `pointer states', and $|0\ra_B$ is the so-called `ready state' of the measurement apparatus before measuring. The reading out of the measurement apparatus can then be described as a measurement performed on the apparatus in the `pointer basis':
\begin{align}
\pr^{\sum_{j=0}^{d-1} c_j |j\ra_A |j\ra_B} \lt( [i]^B \rt).
\end{align}
Note that these probabilities equal the probabilities of the original measurement on system $A$, as they should, because both measurements are actually two different descriptions of a single measurement:
\begin{align}
\pr^{|\psi\ra_A} \lt( \hat{E}^A_i \rt)  =  \pr^{\sum_{j=0}^{d-1} c_j |j\ra_A |j\ra_B} \lt( [i]^B \rt).\label{cj}
\end{align}
Since these probabilities are equal by definition, the same relation holds when considering $\lam$-probabilities:
\begin{align}
\pr_\lam^{|\psi\ra_A} \lt( \hat{E}^A_i \rt)_{\hat{O}^A}  =  \pr_\lam^{\sum_{j=0}^{d-1} c_j |j\ra_A |j\ra_B} \lt( [i]^B \rt)_{\hat{O}^B},\label{cj2}
\end{align}
where $\hat{O}^B$ is an observable that includes the pointer states as eigenstates. Now, to the probability on the right-hand side of the above equation the result of Section \ref{SecAny} can be applied. Therefore, for any measurement of the first kind, the $\lam$-probabilities are trivial:\footnote{More generally, the measurement on the apparatus may involve one or more multidimensional projectors on $\mathcal{H}_B$ that each have multiple pointer states as eigenstates. If this is the case, this corresponds to a degenerate measurement on system $A$ where, for each such projector on $\mathcal{H}_B$, the projectors $\{\hat{E}^A_j\}$ corresponding to these pointer states are summed to obtain the projector on $\mathcal{H}_A$ corresponding to the projector on $\mathcal{H}_B$ (like $\hat{E}^A_i$ corresponds to $[i]^B$ in \eqref{cj} and \eqref{cj2}). The result of Section \ref{SecAny}, also holding for degenerate measurements, still applies. The same considerations apply to the measurements considered in the next two subsections, where in Section \ref{POVMmeas} not projectors, but positive operators $\{\hat{F}^A_j\}$ are summed.}
\begin{align}
\pr_\lam^{|\psi\ra_A} \lt( \hat{E}^A_i \rt) \iae \pr^{|\psi\ra_A} \lt( \hat{E}^A_i \rt) = c_i^2.\label{FinalResult}
\end{align}

\subsection{Measurements of the second kind}
In measurements of the second kind, the coupling between system and measurement apparatus is more general. When a measurement of the first kind is performed, a system initially in an eigenstate of the observable remains in that same eigenstate after the measurement. In contrast, when a measurement of the second kind is performed, such a system may end up in a different state. Therefore, the coupling is as follows:
\begin{align}
|\psi\ra_A|0\ra_B = \sum_{j=0}^{d-1} \hat{E}^A_j |\psi\ra_A |0\ra_B \mapsto \sum_{j=0}^{d-1} c_j |j\ra_A |j\ra_{B}\label{skcoup}
\end{align}
where $c_j$ is as in \eqref{cj}, but now the states $|j\ra_A$ can be anything and are in general not mutually orthogonal. This prevents the direct application of the result of Section \ref{SecAny} to the right-hand side of \eqref{skcoup}. However, the measurement apparatus can generally be decomposed into subsystems which get entangled with each other during the measurement (for example, different particles of a pointer). Or, the measurement apparatus may interact with another system before any interaction with the experimenter takes place (for example, a computer recording the outcome). In this case, we can write $\mathcal{H}_B \equiv \mathcal{H}_{B_1} \otimes \mathcal{H}_{B_2}$, and we have the coupling
\begin{align}
|\psi\ra_A |0\ra_{B_1} |0\ra_{B_2} = \sum_{j=0}^{d-1} \hat{E}^A_j |\psi\ra_A |0\ra_{B_1} |0\ra_{B_2} \mapsto \sum_{j=0}^{d-1} c_j |j\ra_A |j\ra_{B_1}  |0\ra_{B_2}\mapsto \sum_{j=0}^{d-1} c_j |j\ra_A |j\ra_{B_1} |j\ra_{B_2}.\label{b1b2map}
\end{align}
Here, the states $|j\ra_{B_1}$ are mutually orthogonal, as are the states $|j\ra_{B_2}$. Writing $|j\ra_{AB_1} \equiv |j\ra_A |j\ra_{B_1}$, the $\lam$-probabilities of the measurement are now given by
\begin{align}
\pr_\lam^{\sum_{j=0}^{d-1} c_j |j\ra_{AB_1} |j\ra_{B_2}} \lt( [i]^{B_2} \rt)_{\hat{O}^{B_2}}
\end{align}
and, noting that the states $|j\ra_{AB_1}$ are mutually orthogonal, we see that again the result of Section \ref{SecAny} can be applied to it, resulting again in \eqref{FinalResult}. 

\subsection{POVM measurements}\label{POVMmeas}
The result can also be extended to POVM measurements. In the case of such a measurement, the coupling is even more general. It is characterized by a complete set of positive operators $\{\hat{F}^A_j\}_{j=0}^{d-1}$, which do not have to be projection operators. In this case the coupling to the measurement apparatus is as follows:
\begin{align}
|\psi\ra_A |0\ra_B \mapsto \sum_{j=0}^{d-1} \hat{M}^A_j |\psi\ra_A |j\ra_B = \sum_{j=0}^{d-1} c_j |j\ra_A |j\ra_B,
\end{align}
where
\begin{align}
|j\ra_A := \frac{\hat{M}^A_j|\psi\ra_A}{\lt\|\hat{M}^A_j|\psi\ra_A\rt\|} \quad \mbox{, and } \quad c_j := |\la j|\psi\ra_A| = \lt\|\hat{M}^A_j|\psi\ra_A\rt\|,
\end{align}
and the $\hat{M}^A_j$ are operators such that $(\hat{M}^A_j)^\dagger \hat{M}^A_j = \hat{F}^A_j$.

As in the case of measurements of the second kind, the measurement apparatus can generally be decomposed into two subsystems which get entangled during the measurement process:
\begin{align}
|\psi\ra_A |0\ra_{B_1} |0\ra_{B_2} = \sum_{j=0}^{d-1} \hat{M}^A_j |\psi\ra_A |0\ra_{B_1} |0\ra_{B_2} \mapsto \sum_{j=0}^{d-1} c_j |j\ra_A |j\ra_{B_1} |0\ra_{B_2} \mapsto \sum_{j=0}^{d-1} c_i |j\ra_A |j\ra_{B_1} |j\ra_{B_2}.\label{b1b2map}
\end{align}
Again, by writing $|j\ra_{AB_1} \equiv |j\ra_A |j\ra_{B_1}$ we end up with $\lam$-probabilities
\begin{align}
\pr_\lam^{\sum_{j=0}^{d-1} c_j |j\ra_{AB_1} |j\ra_{B_2}} \lt( [i]^{B_2} \rt)_{\hat{O}^{B_2}}
\end{align}
to which the result of Section \ref{SecAny} can be applied, so that we get
\begin{align}
\pr_\lam^{|\psi\ra_A} \lt( \hat{F}^A_i \rt) \iae \pr^{|\psi\ra_A} \lt( \hat{F}^A_i \rt) = c_i^2.
\end{align}
%
%
\section{Discussion}\label{SecDiscussion}
\noindent We summarize the main differences between the deduction presented in this article and C\&R's derivation:
\begin{itemize}
\item As mentioned in the Introduction, C\&R define their `Freedom of Choice' assumption in such a way that they can deduce `no-signalling', which is similar to ParInd, from it (Section VII.A).\footnote{Sections denoted using Roman numerals are sections in \citet{Colb2012}. Sections denoted using Arabic numerals are sections in this article.} Instead, we explicitly assume ParInd, whereby there entire issue of deducing ParInd from a freedom assumption becomes irrelevant.
\item The proof in Section \ref{SecBell} is considerably simplified compared with that of C\&R (Section VII.B).
\item The result of Section \ref{SecDDim} is not present in C\&R's work. Instead, C\&R presume that the result of Section \ref{SecBell} can be extended to maximally entangled states of Schmidt number $2^n$, by considering $n$ copies of a system in a Bell state (Section VII.C). Note that results similar to that of Section \ref{SecDDim} have been derived in \citet{Leif2014}, \citet{Ghir2012} and \citet{Barr2006}.
\item Before generalizing to states with arbitrary Schmidt coefficients in Section \ref{SecAny}, we consider states with Schmidt coefficients that are square roots of rational numbers in Section \ref{SecSqrRat}. C\&R did not perform this intermediate step, and only considered transforming the state approximately to a maximally entangled state with Schmidt number $2^n$. Then, they applied the result for Bell states (Section \ref{SecBell}), without justifying how this result can be applied to approximate states (Section VII.C). Actually, filling this gap is problematic, and therefore in this article a whole different approach, not present in C\&R's work, is introduced to prove the result, including defining the correlation measures for approximate states  $I_{N,n}$ and $I_{J_l,N,n,l}$ and proving Lemma \ref{pLemma} in Appendix \ref{AppLemma}.
\item In contrast to C\&R (Section IV.B), we do not use random variables to represent measurement settings. In our opinion, a free choice is not something that is best described using random variables; any pattern can be chosen for the measurement settings, and a random variable with a well-defined probability distribution does not seem to be the right mathematical object to represent such a free choice \citep[see also][]{Butt1992}.
\end{itemize}

Although Theorem 1 is presented as an impossibility theorem for hidden variable theories satisfying ParInd, it also applies to some theories that violate it. A hidden variable theory violating ParInd may satisfy ParInd after partly integrating over the hidden variables. If such a theory is not trivial at the level of the remaining hidden variables, i.e. measurement outcomes depend on them, then the theory is also shown to be incompatible with QM by the theorem. An example is the class of crypto-nonlocal theories considered by \citet{Legg2003}. Leggett introduces hidden variables $\mathbf{u}$ and $\mathbf{v}$, representing definite photon polarizations, and an additional hidden variable $\lam$. When considering the level where all three hidden variables are included, ParInd is violated, so our theorem does not seem to apply. However, after averaging over $\lam$, ParInd is satisfied, while the values of the remaining hidden variables $\mathbf{u}$ and $\mathbf{v}$ still give information about measurement outcomes: if a photon is measured in the direction $\mathbf{u}$, the outcome is always $+1$. Therefore, Theorem 1 can be applied at this level and thereby Leggett's class of theories is shown to be incompatible with QM. Actually, the result of Section \ref{SecBell} is sufficient to rule out Leggett's models, as shown in \cite{Bran2008}.

One might worry about the fact that from Section \ref{SecSqrRat} onwards, we have made use of extra systems in special embezzling states, and complex measurements performed on those systems. The objection can be raised that not every unitary operation can be implemented in practice, and that not every self-adjoint operator corresponds to a real measurement. There does not seem to be a reason why the states and measurements used are impossible in principle.\footnote{Note that C\&R themselves are working in the field of quantum computing and information, where it is usually assumed that any unitary gate can in principle be implemented and any measurement can in principle be performed.} Therefore, it seems contrived to block the derivation of the result by rejecting the use of certain states and measurements. However, when applying the argument of, for example, Section \ref{SecBell} to a bipartite system consisting of a measurement apparatus $B$ and another system $A$, as is essentially done in Section \ref{SecAnyProj}, the derivation does include measurements with observables having eigenvectors like $1/\!\sqrt2 (|1\ra_B + |2\ra_B)$, where $|1\ra_B$ and $|2\ra_B$ are distinct (macroscopic) pointer states of the measurement apparatus. Such measurements are special in the sense that they can transform an apparatus from the state $|1\ra_B$ into the state $|2\ra_B$. These are also the kind of measurements that can detect whether the measurement apparatus has collapsed, and such measurements are practically very hard, if not impossible, to perform. They are similar to what \citet{Barr1999} calls \emph{A-measurements}, which can empirically distinguish collapse from no-collapse versions of quantum mechanics. Such measurements have to date not be performed, and one might be skeptical about the use of such measurements in the derivation. This worry might be reduced by considering $B$ not to be the whole measurement apparatus, but only a small part of it, while still getting entangled with system $A$ during the measurement process.

As mentioned in the Introduction, C\&R have used their claim to argue that a system's wave function is in a one-to-one correspondence with its ontic state. Roughly, the argument is as follows. C\&R distinguish ontic states on the basis of outcome probabilities: if two states predict the same outcome probabilities for all possible measurements, then the states are considered equal. Now, if two systems in equal quantum states were in different ontic states, a variable could be introduced which represents the ontic state of the system. Furthermore, outcome probabilities would depend on this variable, contradicting C\&R's claim. Therefore, equal quantum states imply equal ontic states, and the outcome probabilities for a quantum state equal those for the corresponding ontic state.

Now suppose that two systems are in equal ontic states, but in different quantum states. Following the above, outcome probabilities for both quantum states would equal those for the ontic state. But this contradicts the fact that in quantum mechanics, different quantum states imply different outcome probabilities for some measurements. Therefore, equal quantum states imply equal ontic states. It follows that there is a one-to-one correspondence between quantum states and ontic states.

Of course, this result is based on C\&R's strong claim that no non-trivial hidden variable theory is compatible with quantum mechanics. If the theorem as presented in this article is used, a weaker result follows: there is a one-to-one correspondence between quantum states and ontic states, \emph{if, at the level of the ontic states, ParInd is satisfied}.

\section{Acknowledgements}
\noindent I would like to thank F.A. Muller, Klaas Landsman, Dennis Dieks, Roger Colbeck and Guido Bacciagaluppi for valuable discussions and corrections. This work is part of the research programme `The Structure of Reality and the Reality of Structure', which is (partly) financed by the Netherlands Organisation for Scientific Research (NWO).

\appendix
\section{Calculation of correlation measures}
\noindent In this appendix it is shown that, similar to $\lim_{N \to \infty} I_N = 0$, we also have $\lim_{N \to \infty} \lim_{n \to \infty} I_{N,n} = 0$ and $\lim_{N \to \infty} \lim_{l \to \infty} \lim_{n \to \infty} I_{J_l, N, n, l} = 0$.

\subsection{$I_{N,n}$}\label{SubINn}
Recall the definitions from Section \ref{SecSqrRat}:
\begin{align}
|\chi_n\ra	&:= |\tau_n\ra_{A''B''} \otimes \sum_{i=0}^{d-1} \lt( \sqrt{\frac{m_i}r}|i\ra_A|i\ra_B\sum_{j=0}^{m_i-1} \frac1{\sqrt{m_i}}|j\ra_{A'}|j\ra_{B'}\rt),\notag\\
I_{N,n} &:= \sum_{|a-b|=1} \pr^{|\Psi_n\ra} \lt(A_{N,n,a} \ne B_{N,n,b}\rt)\mbox{, where}\notag\\
|\Psi_n\ra &:= |\tau_n\ra_{A''B''} \otimes |0\ra_{A'}|0\ra_{B'} \otimes |\phi_d\ra_{AB},\notag\\
\hat{A}_{N,n,a} &:= \lt(U^{AA'A''}_n\rt)^{-1} \lt( {\mathbb{I}}^{A''} \otimes \hat{O}^{AA'}_{a\pi/2N}\rt)\lt(U^{AA'A''}_n\rt),\quad a \in \mathcal{A}_N,\notag\\
\hat{B}_{N,n,b} &:= \lt(U^{BB'B''}_n\rt)^{-1} \lt( {\mathbb{I}}^{B''} \otimes \hat{O}^{BB'}_{b\pi/2N}\rt)\lt(U^{BB'B''}_n\rt),\quad b \in \mathcal{B}_N.
\end{align}
Instead of including the unitary operators $U^{AA'A''}_n, U^{BB'B''}_n$ in the definition of $\hat{A}_{N,n,a}$ and $\hat{B}_{N,n,b}$, they can also be attached to the state $|\Psi_n\ra$. In this case, the quantum probabilities remain the same. Define
\begin{align}
\hat{\overline{A}}_{N,a} := {\mathbb{I}}^{A''} \otimes \hat{O}^{AA'}_{a\pi/2N}, \quad  a \in \mathcal{A}_N;\notag\\
\hat{\overline{B}}_{N,b} := {\mathbb{I}}^{B''} \otimes \hat{O}^{BB'}_{b\pi/2N}, \quad  b \in \mathcal{B}_N.
\end{align}
Note that these are just $\hat{A}_{N,n,a}$ and $\hat{B}_{N,n,b}$ \eqref{manydefs} without the unitary operators. Then,
\begin{align}
I_{N,n} = \sum_{|a-b|=1} \pr^{\lt(U^{AA'A''}_n \otimes U^{BB'B''}_n\rt)|\Psi_n\ra} \lt(\overline{A}_{N,a} \ne \overline{B}_{N,b} \rt).
\end{align}
For the state $|\chi_n\ra$, which is close to $\lt(U^{AA'A''}_n \otimes U^{BB'B''}_n\rt)|\Psi_n\ra$ for large $n$, a calculation of quantum probabilities yields for any $a$, $b$ with $|a-b| = 1$:
\begin{align}
\pr^{|\chi_n\ra} \lt(\overline{A}_{N,a} \ne \overline{B}_{N,b} \rt) = \frac2d\sin^2 \lt( \frac{\pi}{4N} \rt).\label{PABQMProb}
\end{align}
From the property of the trace distance \eqref{StateProbTD} , we have
\begin{align}
\lt| \pr^{\lt(U^{AA'A''}_n \otimes U^{BB'B''}_n\rt)|\Psi_n\ra} \lt(\overline{A}_{N,a} \ne \overline{B}_{N,b} \rt) - \pr^{|\chi_n\ra} \lt(\overline{A}_{N,a} \ne \overline{B}_{N,b} \rt)\rt|\notag\\
\le D\lt(\lt(U^{AA'A''}_n \otimes U^{BB'B''}_n\rt)|\Psi_n\ra, |\chi_n\ra\rt).\label{INnDeriv1}
\end{align}
Taking the limit $n \to \infty$ on the right-hand side yields 0 by \eqref{Eq553}, and therefore, also using \eqref{PABQMProb}, for $|a-b| = 1$,
\begin{align}
&\lim_{n \to \infty} \lt| \pr^{\lt(U^{AA'A''}_n \otimes U^{BB'B''}_n\rt)|\Psi_n\ra} \lt(\overline{A}_{N,a} \ne \overline{B}_{N,b} \rt) - \frac2d\sin^2 \lt( \frac{\pi}{4N} \rt)\rt| = 0\notag\\
\Rightarrow \quad &\lim_{n \to \infty} \pr^{\lt(U^{AA'A''}_n \otimes U^{BB'B''}_n\rt)|\Psi_n\ra} \lt(\overline{A}_{N,a} \ne \overline{B}_{N,b} \rt) = \frac2d\sin^2 \lt( \frac{\pi}{4N} \rt).
\end{align}
Finally, because $I_{N,n}$ is the sum of $2N$ such probabilities,
\begin{align}
\lim_{n \to \infty} I_{N,n} = 2N\frac2d \sin^2 \lt( \frac{\pi}{4N} \rt) \le \frac{\pi^2}{4Nd} \quad \Rightarrow \quad \lim_{N \to \infty} \lim_{n \to \infty} I_{N,n} = 0.\label{INnDeriv2}
\end{align}
\subsection{$I_{J_l, N, n, l}$}\label{SubIJL}
Recall the definitions from Section \ref{SecAny}:
\begin{align}
I_{J_l, N, n, l} &:= \sum_{|a-b|=1} \pr^{|\Psi_n\ra}\lt(A_{J_l, N, n, l, a} \ne B_{J_l, N, n, l, b}\rt)\mbox{, where}\notag\\
|\Psi_n\ra &:= |\tau_n\ra_{A''B''} \otimes |0\ra_{A'} |0\ra_{B'} \otimes \sum_{i=0}^{d-1} c_{i}|i\ra_A|i\ra_B,\notag\\
\hat{A}_{J_l, N, n, l, a} &:= \lt(U^{AA'A''}_{n,l}\rt)^{-1} \lt({\mathbb{I}}^{A''} \otimes \hat{O}^{AA;}_{J_l, a\pi/2N}\rt) \lt(U^{AA'A''}_{n,l}\rt), \quad a \in \mathcal{A}_N\!\setminus\!\{2N\},\notag\\
\hat{B}_{J_l, N, n, l, b} &:= \lt(U^{BB'B''}_{n,l}\rt)^{-1} \lt({\mathbb{I}}^{B''} \otimes \hat{O}^{BB;}_{J_l, b\pi/2N}\rt) \lt(U^{BB'B''}_{n,l}\rt), \quad b \in \mathcal{B}_N\!\setminus\!\{2N\}.
\end{align}
As in Appendix \ref{SubINn}, we move the unitary operators from the observables to the state. Define
\begin{align}
|\chi_{n,l}\ra &:= |\tau_n\ra_{A''B''} \otimes \sum_{i=0}^{d-1} \sum_{j=0}^{m_{i,l} -1} \frac{c_i}{\sqrt{m_{i,l}}}|i,j\ra_{AA'}|i,j\ra_{BB'};\notag\\
|\Phi_{n,l}\ra &:= |\tau_n\ra_{A''B''} \otimes \sum_{i=0}^{d-1} \sum_{j=0}^{m_{i,l} -1} \frac1{\sqrt{r_l}} |i,j\ra_{AA'}|i,j\ra_{BB'};\notag\\
\hat{\overline{A}}_{J_l,N,a} &:= {\mathbb{I}}^{A''} \otimes \hat{O}^{AA'}_{J_l, a\pi/2N}, \quad a \in \mathcal{A}_N \!\setminus\! \{2N\};\notag\\
\hat{\overline{B}}_{J_l,N,b} &:= {\mathbb{I}}^{B''} \otimes \hat{O}^{BB'}_{J_l, b\pi/2N}, \quad b \in \mathcal{B}_N; \notag\\
\hat{\overline{A}}_{J_l,N,2N} &:= - \hat{\overline{A}}_{J_l,N,0}.
\end{align}
Then,
\begin{align}
I_{J_l,N, n, l} = \sum_{|a-b|=1} \pr^{\lt(U^{AA'A''}_{n,l} \otimes U^{BB'B''}_{n,l}\rt)|\Psi_n\ra} \lt(\overline{A}_{J_l,N,a} \ne \overline{B}_{J_l,N,b} \rt).
\end{align}
Quantum mechanics tells us that for the state $|\Phi_{n,l}\ra$, for any $a$, $b$ with $|a-b| = 1$,%
\begin{align}
\pr^{|\Phi_{n,l}\ra} \lt(\overline{A}_{J_l,N,a} \ne \overline{B}_{J_l,N,b} \rt) = \sin^2 \lt( \frac{\pi}{4N} \rt).\label{ABJlQM}
\end{align}
Also, in the limit $l \to \infty$, the states $|\chi_{n,l}\ra$ and $|\Phi_{n,l}\ra$ approach each other:
\begin{align}
1 \ge \mathcal{F}(|\chi_{n,l}\ra, |\Phi_{n,l}\ra) &= |\la\chi_{n,l}|\Phi_{n,l}\ra| = \sum_{i=0}^{d-1} \sum_{j=0}^{m_{i,l} - 1} m_{i,l} \frac{c_i}{\sqrt{r_l \cdot m_{i,l}}} = \sum_{i=0}^{d-1} \sqrt{\frac{m_{i,l}}{r_l}}\cdot c_i \notag\\
&= \sum_{i=0}^{d-1} c_{i,l} \cdot c_i = 1 + \sum_{i=0}^{d-1} c_i (c_{i,l} - c_i) \ge 1 - \sum_{i=0}^{d-1} c_i |c_{i,l} - c_i|\notag\\
&\Rightarrow \quad \lim_{l \to \infty} \mathcal{F}(|\chi_{n,l}\ra, |\Phi_{n,l}\ra) = 1 \mbox{ and } \lim_{l \to \infty} D\lt(|\chi_{n,l}\ra, |\Phi_{n,l}\ra\rt) = 0,\label{SecondTermZero}
\end{align}
where we used \eqref{Limcil}, as well as the relation between fidelity and trace distance \eqref{RelFidTD}. Using the triangle inequality $| | x - y | - | y - z| | \le | x - z |$ and \eqref{StateProbTD}, we have, omitting $( \overline{A}_{J_l,N,a} \ne \overline{B}_{J_l,N,b} )$ after every $\pr$ to avoid cluttered notation,
\begin{align}
&\lt| \lt| \pr^{\lt(U^{AA'A''}_{n,l} \otimes U^{BB'B''}_{n,l}\rt)|\Psi_n\ra}  - \pr^{|\Phi_{n,l}\ra} \rt| - \lt| \pr^{|\Phi_{n,l}\ra} - \pr^{|\chi_{n,l}\ra} \rt| \rt|\notag\\
&\le \lt| \pr^{\lt(U^{AA'A''}_{n,l} \otimes U^{BB'B''}_{n,l}\rt)|\Psi_n\ra} - \pr^{|\chi_{n,l}\ra} \rt| \le D\lt( \lt(U^{AA'A''}_{n,l} \otimes U^{BB'B''}_{n,l}\rt)|\Psi_n\ra , |\chi_{n,l}\ra \rt).
\end{align}
In the limit $n \to \infty$, the right-hand side vanishes by \eqref{AppDResult}, and therefore
\begin{align}
\lim_{n \to \infty} \lt| \lt| \pr^{\lt(U^{AA'A''}_{n,l} \otimes U^{BB'B''}_{n,l}\rt)|\Psi_n\ra} - \pr^{|\Phi_{n,l}\ra} \rt| - \lt| \pr^{|\Phi_{n,l}\ra} - \pr^{|\chi_{n,l}\ra} \rt| \rt| = 0.
\end{align}
Since $\lt| \pr^{|\Phi_{n,l}\ra} - \pr^{|\chi_{n,l}\ra} \rt|$ is independent of $n$ we have\footnote{The states $|\Phi_{n,l}\ra$ and $|\chi_{n,l}\ra$ only depend on $n$ through the embezzling states $|\tau_n\ra$ of the subsystems $A''$ and $B''$, but the measurements are performed on the subsystems $AA'$ and $BB'$.}
\begin{align}
\lim_{n \to \infty} \lt| \pr^{\lt(U^{AA'A''}_{n,l} \otimes U^{BB'B''}_{n,l}\rt)|\Psi_n\ra} - \pr^{|\Phi_{n,l}\ra} \rt| = \lt| \pr^{|\Phi_{n,l}\ra} - \pr^{|\chi_{n,l}\ra} \rt| \le D\lt(|\chi_{n,l}\ra, |\Phi_{n,l}\ra\rt),
\end{align}
where we also used \eqref{RelFidTD} again. By \eqref{SecondTermZero}, the right-hand side vanishes in the limit $l \to \infty$, and therefore we have, also applying \eqref{ABJlQM}, for any $a$, $b$ with $|a-b| = 1$,%
\begin{align}
\lim_{l \to \infty} \lim_{n \to \infty} \lt| \pr^{\lt(U^{AA'A''}_{n,l} \otimes U^{BB'B''}_{n,l}\rt)|\Psi_n\ra} \lt( \overline{A}_{J_l,N,a} \ne \overline{B}_{J_l,N,b} \rt) -  \sin^2 \lt( \frac{\pi}{4N} \rt) \rt| = 0.
\end{align}
The result now follows by following steps similar to \eqref{INnDeriv1}-\eqref{INnDeriv2}.
\begin{align}
&\lim_{l \to \infty} \lim_{n \to \infty} \pr^{\lt(U^{AA'A''}_{n,l} \otimes U^{BB'B''}_{n,l}\rt)|\Psi_n\ra} \lt(\overline{A}_{J_l,N,a} \ne \overline{B}_{J_l,N,b} \rt) = \sin^2 \lt( \frac{\pi}{4N} \rt) \notag\\
&\Rightarrow \lim_{l \to \infty} \lim_{n \to \infty} I_{J_l, N, n, l} = 2N \sin^2 \lt( \frac{\pi}{4N} \rt)\notag\\
&\Rightarrow \lim_{N \to \infty} \lim_{l \to \infty} \lim_{n \to \infty} I_{J_l, N, n, l} = 0.
\end{align}

\section{Lemma 1}\label{AppLemma}
\noindent This lemma is based on the proposition that if we have $r$ numbers, where $r$ is even, and the sum of any half of these numbers is close to $1/2$, then the sum of $j$ of these numbers is close to $j/r$. One might guess that this can be proven by first showing that any individual number is close to $1/r$, and than concluding that the sum of $r/2$ of such numbers are close to $1/2$, but it turns out that in our case this does not work. Instead, we make use of the fact that any sum of numbers can be expressed as a linear combination of sums of $r/2$ numbers. For example, suppose we have the sequence $(p_i)_{i=0}^{9}$, where the sum of any five members of the sequence is within $\epsilon$ from $1/2$. We can express any sum of $p_i$'s as a linear combination of such sums, for example:
\begin{align}
\sum_{i \in \{0,1\}} p_i = \frac15&\lt(\sum_{i \in \{0,1,2,3,4\}}\!\!\!\!\!\! p_i + \!\!\! \sum_{i \in \{0,1,5,6,7\}}\!\!\!\!\!\! p_i + \!\!\!\sum_{i \in \{0,1,2,8,9,2\}}\!\!\!\!\!\! p_i+ \!\!\!\sum_{i \in \{0,1,3,4,5\}}\!\!\!\!\!\! p_i + \!\!\!\sum_{i \in \{1,2,7,8,9\}}\!\!\!\!\!\! p_i \rt.\notag\\
&\lt.- \!\!\!\sum_{i \in \{2,3,4,5,6\}}\!\!\!\!\!\! p_i - \!\!\!\sum_{i \in \{7,8,9,2,3\}}\!\!\!\!\!\! p_i - \!\!\!\sum_{i \in \{4,5,6,7,8\}}\!\!\!\!\!\!p_i \rt).\label{plisum}
\end{align}
Since we add or subtract eight sums that are within $\epsilon$ close of $1/2$, and we divide by $5$, we find that $\sum_{i \in \{1,2\}} p_i$ is within $8\epsilon/5$ of $1/5 = 2/10$. In the proof below, we generalize this result, adding an extra index $\lam$ because we consider not a single sequence but a collection of sequences over which we average using the measure $\mu$.%
\addtocounter{lemma}{-1}
\begin{lemma}\label{pLemma}
Let $\Lambda$ be a measurable space with measure $\mu$, let $r$ be an even, positive integer, and let
\begin{align}
\lt\{\lt(p^\lam_i\rt)_{i=0}^{r-1} \me| \lam \in \Lambda\rt\}
\end{align}
be some collection of sequences satisfying, for all $I \subset I^{\mathrm{ind}} := \mathbb{N}_r$ with $\#I = r/2$,
\begin{align}
\int_\Lambda\mathrm{d}\mu(\lam) \lt|\lt(\sum_{i \in I} p^\lam_i \rt) -1/2 \rt| &< \epsilon.\label{Lem2Ante}
\end{align}
Then for all $J \subset I^{\mathrm{ind}}$
\begin{align}
\int_\Lambda\mathrm{d}\mu(\lam) \lt| \lt(\sum_{i \in J} p^\lam_i \rt) - {\#J}/r \rt| &< 2\epsilon.
\end{align}
\end{lemma}
First the case $\#J \le r/2$ is considered. Let $J \subset I^{\mathrm{ind}}$, $\#J \le r/2$, and $\epsilon > 0$. By \eqref{Lem2Ante} we have, for all $I \subset I^{\mathrm{ind}}$ with $\#I = r/2$,
\begin{align}
\int_\Lambda\mathrm{d}\mu(\lam) \lt|\lt(\sum_{i \in I} p^\lam_i \rt) -1/2 \rt| < \epsilon.\label{Cstar}
\end{align}
Define the sequence $(f_l)_{l=0}^{r-\#J-1}$ containing the elements of $I^{\mathrm{ind}} \!\setminus\! J$ in arbitrary order. Also define
\begin{align}
K_a &:= J \cup \lt\{f_{\overline{ax}}, f_{\overline{ax +1}}, \dots, f_{\overline{ax + (x - 1)}}\rt\}\notag\\
&\mbox{where } x := r/2 - \#J, \quad \overline{y} := y\;\mathrm{mod}\;(r - \#J), \quad a \in \mathbb{N}_{r/2};\notag\\
L_b &= \lt\{f_{\overline{b(r/2)}}, f_{\overline{b(r/2) + 1}}, \dots, f_{\overline{b(r/2) + (r/2 - 1)}}\rt\}, \quad b \in \mathbb{N}_{r/2 - \#J}.
\end{align}
Note that for all $a$ and $b$, $\#K_a = \#L_b = r/2$. The $K_a$ are index sets for the sums of $p$'s to be \emph{added}, like the index sets $\{0,1,2,3,4\}, \{0,1,5,6,7\}, \dots$ in \eqref{plisum}. The $L_b$ are index sets for the sums of $p$'s to be \emph{subtracted}, like the index sets $\{2,3,4,5,6\}, \{7,8,9,2,3\}, \dots$ in \eqref{plisum}.

Define
\begin{align}
R^\lam_a := \sum_{k \in K_{a}} p^\lam_k \; \mbox { and } \; T^\lam_b := \sum_{l \in L_b} p^\lam_l.
\end{align}
Then
\begin{align}
\sum_{i \in J}p^\lam_i = \frac{\sum_{a=0}^{r/2 - 1}R^\lam_a - \sum_{b=0}^{r/2 - \#J - 1}T^\lam_b}{r/2},\label{C7}
\end{align}
and
\begin{align}
\int \mathrm{d}\mu(\lam)\lt|\sum_{i \in J}p^\lam_i - {\#J}/r\rt|
&= \frac1{r/2}\int_\Lambda \mathrm{d}\mu(\lam) \lt| \sum_{a=0}^{r/2 - 1}R^\lam_a - \sum_{b=1}^{r/2-\#J}T^\lam_b - {\#J}/2 \rt |\notag\\
&=\int_\Lambda\mathrm{d}\mu(\lam)\frac1{r/2} \lt| \sum_{a=0}^{r/2 - 1}\lt(R^\lam_a - 1/2\rt) - \sum_{b=0}^{r/2 - \#J - 1}\lt(T^\lam_b - 1/2\rt) \rt|\notag\\
&\le \int_\Lambda\mathrm{d}\mu(\lam)\frac1{r/2} \lt( \sum_{a=0}^{r/2 - 1}\lt|R^\lam_a - 1/2\rt| + \sum_{b=0}^{r/2 - \#J - 1}\lt|T^\lam_b - 1/2\rt| \rt)\notag\\
&=\frac1{r/2} \lt( \sum_{a=0}^{r/2 - 1}\int_\Lambda\mathrm{d}\mu(\lam)\lt|R^\lam_a - 1/2\rt| + \sum_{b=0}^{r/2 - \#J - 1}\int_\Lambda\mathrm{d}\mu(\lam)\lt|T^\lam_b - 1/2\rt| \rt)\notag\\
&\le \frac{r - \#J}{r/2}\epsilon < 2\epsilon,\notag\\
\end{align}
where we used \eqref{C7} and \eqref{Cstar}. For the case $\#J > r/2$, define $K := I^{\mathrm{ind}}\!\setminus\! J$. Then $\#K = r - \#J < r/2$ and
\begin{align}
\int_\Lambda \mathrm{d}\mu(\lam)\lt|\sum_{i \in J}p^\lam_i - {\#J}/r\rt| &= \int_\Lambda \mathrm{d}\mu(\lam)\lt|\lt(1-\sum_{i \in K}p^\lam_i\rt) - \lt(1 - {\#K}/r\rt)\rt|\notag\\
&= \int_\Lambda \mathrm{d}\mu(\lam)\lt|\sum_{i \in K}p^\lam_i - {\#K}/r\rt| < 2\epsilon.
\end{align}
Concluding, for any $J \subset I^\mathrm{ind}$,
\begin{align}
\int_\Lambda\mathrm{d}\mu(\lam) \lt|\sum_{i \in J} p^\lam_i - {\#J}/r \rt| < 2\epsilon .
\end{align}
\section{Embezzlement}\label{AppEmb}
\noindent Recall the definitions:
\begin{align}
|\chi_{n,l}\ra	&:= |\tau_n\ra_{A''B''} \otimes \sum_{i=0}^{d-1} \sum_{j=0}^{m_{i,l} -1} \frac{c_i}{\sqrt{m_{i,l}}}|i,j\ra_{AA'}|i,j\ra_{BB'};\notag\\
			&= \sum_{i=0}^{d-1} \sum_{k=0}^{n-1} \sum_{j=0}^{m_{i,l} -1} \frac{c_{i}}{\sqrt{C_n m_{i,l}(k+1)}}|i\ra_A|i\ra_B|j\ra_{A'}|j\ra_{B'}|k\ra_{A''}|k\ra_{B''}.\notag\\
|\Psi_n\ra &:= |\tau_n\ra_{A''B''} \otimes |0\ra_{A'}|0\ra_{B'} \otimes \sum_{i=0}^{d-1} c_i |i\ra_A |i\ra_B;\notag\\
|\tau_n\ra &:= \frac1{\sqrt{C_n}}\sum_{k=0}^{n-1}\frac1{\sqrt{k+1}}|k\ra \otimes |k\ra;\notag\\
C_n &:= \sum_{k=0}^{n-1} \frac1{k+1};\notag\\
r_l &:= \mathrm{LCD}\lt(\lt\{c_{i,l}^2\rt\}_{i=0}^{d-1} \cup 1/2\rt)\notag\\
m_{i,l} &:= r_l c^2_{i,l}
\end{align}
Note that the derivation in this section also holds without the subscripts $l$. In that case the definition of $r_l$ is replaced by
\begin{align}
r &:= \mathrm{LCD}\lt(\lt\{c_{i}^2\rt\}_{i=0}^{d-1}\rt);\notag\\
\end{align}

Define $\lfloor x \rfloor$ as the \emph{floor} of $x$: the greatest integer not greater than $x$, and $\lceil x \rceil$ as the \emph{ceiling} of $x$: the smallest integer not smaller than $x$. Define the unitary operator $U^{AA'A''}_{n,l}$ with, for $k \in \mathbb{N}_n$ and $i \in \mathbb{N}_d$:
\begin{align}
U^{AA'A''}_{n,l} \lt( |k\ra_{A''} |0\ra_{A'} |i\ra_A \rt) = \lt| \lt\lfloor {k}/{m_{i,l}} \rt\rfloor \rt\ra_{A''} |k\;\mathrm{mod}\;m_{i,l}\ra_{A'} |i\ra_A,\label{DefU}
\end{align}
and an arbitrary action on the other basis vectors. $U^{BB'B''}_{n,l}$ is defined similarly. Then,
\begin{align}
&\lt( U^{AA'A''}_{n,l} \otimes U^{BB'B''}_{n,l} \rt) |\Psi_n\ra\notag\\
&= \sum_{i=0}^{d-1} \sum_{k=0}^{n-1} \frac{c_i}{\sqrt{C_n(k+1)}}|i\ra_A |i\ra_B |k\;\mathrm{mod}\;m_{i,l}\ra_{A'}|k\;\mathrm{mod}\;m_{i,l}\ra_{B'} \lt| \lt\lfloor k/m_{i,l} \rt\rfloor \rt\ra_{A''} \lt| \lt\lfloor k/m_{i,l} \rt\rfloor \rt\ra_{B''}.
\end{align}
Calculating the fidelity:
\begin{align}
\mathcal{F} \lt( |\chi_{n,l}\ra, \lt( U^{AA'A''}_{n,l} \otimes U^{BB'B''}_{n,l} \rt) |\Psi_n\ra \rt) &= |\la\chi_{n,l}|\lt( U^{AA'A''}_{n,l} \otimes U^{BB'B''}_{n,l} \rt) |\Psi_n\ra| \notag\\
&= \sum_{i=0}^{d-1} \sum_{k=0}^{n-1} \frac{c_{i}}{\sqrt{C_n(k+1)}} \frac{c_{i}}{\sqrt{C_n m_{i,l} \lt( \lt \lfloor k/m_{i,l} \rt \rfloor + 1 \rt)}} \notag\\
&= \sum_{i=0}^{d-1} \sum_{k=0}^{n-1} \frac{c^2_{i}}{C_n\sqrt{m_{i,l}(k+1) \lt( \lt\lfloor k/m_{i,l} \rt\rfloor + 1 \rt)}}.
\end{align}
Now, using $k + 1 \le m_{i,l}^2 \lceil (k + 1)/m_{i,l} \rceil$ and $\lfloor k/m_{i,l} \rfloor + 1 \le \lceil (k + 1)/m_{i,l} \rceil$,
\begin{align}
&\sum_{k=0}^{n-1}\frac1{\sqrt{m_{i,l}(k+1) \lt( \lt\lfloor k/m_{i,l} \rt\rfloor + 1 \rt)}} \notag\\
&\ge \sum_{k=0}^{n-1}\frac1{\sqrt{m^2_{i,l} \lt\lceil {(k+1)}/{m_{i,l}} \rt\rceil \lt( \lt\lfloor k/m_{i,l} \rt\rfloor + 1 \rt)}} \ge \sum_{k=0}^{n-1} \frac1{m_{i,l} \lt\lceil {(k+1)}/{m_{i,l}} \rt\rceil}\notag\\
&= \sum_{k=0}^{m_{i,l}-1} \frac1{m_{i,l}} + \sum_{k=m_{i,l}}^{2m_{i,l}-1} \frac1{2m_{i,l}} + \dots + \sum_{k=\lfloor n/m_{i,l}\rfloor m_{i,l}}^{n-1} \frac1{(\lfloor n/m_{i,l}\rfloor + 1) m_{i,l}}\notag\\
&= \lt( \sum_{k' =1}^{\lt\lceil {n}/{m_{i,l}} \rt\rceil} \frac1{k'} \rt) + \lt( \frac{n}{m_{i,l}} - \lt\lfloor \frac{n}{m_{i,l}} \rt\rfloor \rt) \frac1{\lt\lfloor {n}/{m_{i,l}} \rt\rfloor +1}\notag\\
&= \mathcal{Z}\lt(\frac{n}{m_i,l}\rt), \mbox{where } \mathcal{Z}(y) := \lt( \sum_{k = 1}^{\lceil y \rceil} \frac1k \rt) + \lt( y - \lfloor y \rfloor \rt) \frac1{\lfloor y \rfloor + 1}
\end{align}
\begin{figure}
\begin{center}
\includegraphics[width=12cm]{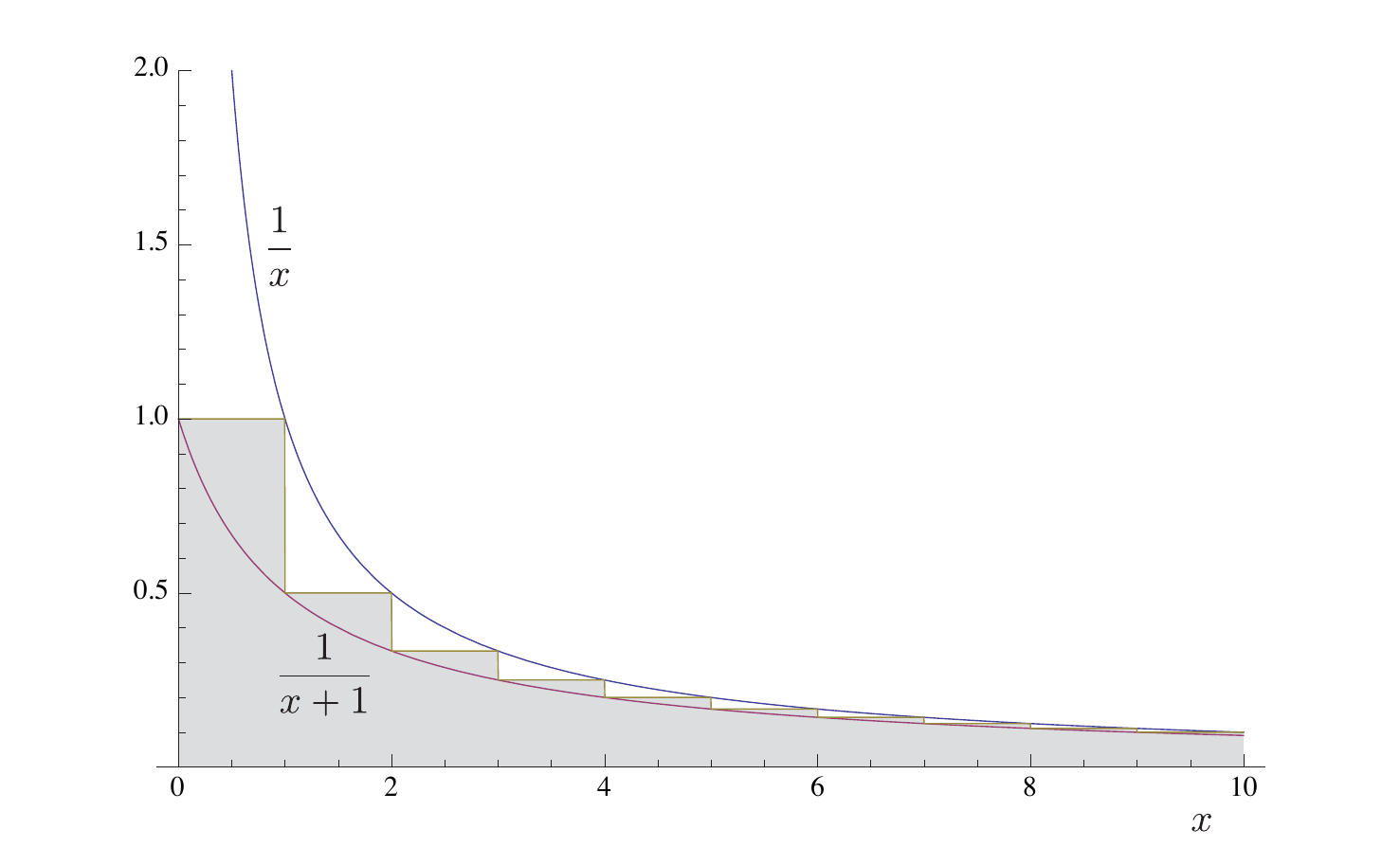}\label{FiggHN}
\caption{Using the two plotted graphs, bounds for the grey area can be derived.}\label{FiggHN}
\end{center}
\end{figure}
Note that for integer values of $y$, $\mathcal{Z}(y) = \sum_{k = 1}^{y} 1/k = C_y$. For non-integer values, $\mathcal{Z}(y)$ equals the grey area in Figure \ref{FiggHN} between $x=0$ and $x=y$. From this we can read off the inequalities:
\begin{align}
\mathcal{Z}(y) &\le 1 + \int_1^y \frac1{x}\mathrm{d}x = 1+\ln y\\
\mathcal{Z}(y) &\ge \int_0^y \frac1{x+1}\mathrm{d}x = \ln (y+1)\\
\mathcal{Z}(y_2) - \mathcal{Z}(y_1) &\le \int_{y_1}^{y_2} \frac1{x}\mathrm{d}x = \ln y_2 - \ln y_1.
\end{align}
Now, noting that
\begin{align}
\mathcal{F} \lt( |\chi_{n,l}\ra, \lt( U^{AA'A''}_{n,l} \otimes U^{BB'B''}_{n,l} \rt) |\Psi_n\ra \rt) = \sum_{i=0}^{d-1} c_i^2 \frac{\mathcal{Z}\lt(\frac{n}{m_{i,l}} \rt)}{\mathcal{Z}(n)}
\end{align}
and
\begin{align}
\frac{\mathcal{Z}\lt(\frac{n}{m_{i,l}} \rt)}{\mathcal{Z}(n)} = 1 - \frac{\mathcal{Z}(n) - \mathcal{Z}\lt(\frac{n}{m_{i,l}} \rt)}{\mathcal{Z}(n)} &> 1 - \frac{\ln m_{i,l}}{\ln n},
\end{align}
it follows, using \eqref{RelFidTD}, that
\begin{align}
D\lt(|\chi_{n,l}\ra, \lt( U^{AA'A''}_{n,l} \otimes U^{BB'B''}_{n,l} \rt) |\Psi_n\ra\rt) &< \sqrt{1 - \lt(1 - \frac{\ln m_{i,l}}{\ln n} \rt)^2}\notag\\
\Rightarrow \lim_{n \to \infty}  D\lt(|\chi_{n,l}\ra, \lt( U^{AA'A''}_{n,l} \otimes U^{BB'B''}_{n,l} \rt) |\Psi_n\ra\rt) &= 0.\label{AppDResult}
\end{align}

\section{Perfect Correlation}\label{AppPC}
\noindent Quantum mechanics predicts that there is perfect correlation between the outcomes of certain measurements on $A$ and $B$. For example, for a measurement on the state $|\phi_d\ra_{AB} = \sum_{i=0}^{d-1}c_i|i\ra_A|i\ra_B$, when the outcome on one side corresponds to the projector $[i]^A$, the outcome on the other side corresponds to the projector $[i]^B$, and vice versa. Therefore, the probabilities of these outcomes must be equal. This property remains when considering $\lam$-probabilities. Similar properties hold for more complex measurements on $AA'A''$ and $BB'B''$. What follows is a formal proof of this fact. First, recall the definitions
\begin{align}
|\Psi_n\ra &:= |\tau_n\ra_{A''B''} \otimes |0\ra_{A'}|0\ra_{B'} \otimes |\phi_d\ra_{AB},\notag\\
\hat{E}_{(i,j),n,l} &:= \lt(U^{AA'A''}_{n,l}\rt)^{-1} \lt( [i,j]^{AA'} \otimes {\mathbb{I}}^{A''} \rt) \lt(U^{AA'A''}_{n,l} \rt),\notag\\
\hat{F}_{(i,j),n,l} &:= \lt(U^{BB'B''}_{n,l}\rt)^{-1} \lt( [i,j]^{BB'} \otimes {\mathbb{I}}^{B''} \rt) \lt(U^{BB'B''}_{n,l} \rt).
\end{align}

\begin{lemma}[Perfect correlation] The following identities hold:
\begin{align}
\forall I \subset \mathbb{N}_d \!: \pr_\lam^{|\phi_d\ra_{AB}} \lt( \sum_{i \in I} [i]^A \rt) &\iae \sum_{i \in I} \pr_\lam^{|\phi_d\ra_{AB}} \lt( [i]^B \rt),\label{PC0}\\
\forall I_l \subset J^{\mathrm{ind}}_{l} \!: \pr_\lam^{|\Psi_n\ra} \lt(\sum_{(i,j) \in I_l} \hat{E}_{(i,j),n,l}\rt) &\iae \sum_{(i,j) \in I_l} \pr_\lam^{|\Psi_n\ra} \lt(\hat{F}_{(i,j),n,l}\rt)\label{PC3}\\
\sum_{j=0}^{m_{i,l}-1}\pr_\lam^{|\Psi_n\ra} \lt(\hat{E}_{(i,j),n,l}\rt) &\iae \pr_\lam^{|\Psi_n\ra} \lt([i]^B\rt), \mbox{ and }\label{PC1}\\
\sum_{j=0}^{m_{i,l}-1}\pr_\lam^{|\Psi_n\ra} \lt(\hat{F}_{(i,j),n,l}\rt) &\iae \pr_\lam^{|\Psi_n\ra} \lt([i]^A\rt).\label{PC2}
\end{align}
\end{lemma}
\noindent Note that the lemma above and the proof below also hold with the subscripts $l$ removed. In this proof, probabilities are considered for the event that an outcome does \emph{not} correspond to a projector $[i]$. We denote such an event as $\neg [i]$. Also, probabilities are considered for the event that an outcome does not correspond to any projector from a set indexed by $I$. In that case, the event will be written as $\neg \bigvee_{i \in I} [i]$.

First, note that according to the Born rule,
\begin{align}
\sum_{i' \in I} \pr^{|\phi_d\ra_{AB}} \lt( \sum_{i \in I} [i]^A, [i']^B \rt) = \sum_{i' \in I} {}_{AB}\la\phi_d| \sum_{i \in I} [i]^A \otimes [i']^B |\phi_d\ra_{AB},
\end{align}
where $|\phi_d\ra_{AB}$ has the form
\begin{align}
|\phi_d\ra_{AB} = \sum_{i=0}^{d-1} c_i |i\ra_A |i\ra_B.
\end{align}
Using this and applying CompQuant, it follows that
\begin{align}
\pr^{|\phi_d\ra_{AB}} \lt( \sum_{i \in I} [i]^A, \neg \bigvee_{i' \in I} [i']^B \rt)  = 0 \quad \Rightarrow \quad \pr_\lam^{|\phi_d\ra_{AB}} \lt( \sum_{i \in I} [i]^A, \neg \bigvee_{i' \in I} [i']^B \rt)  \iae 0, \notag\\
\sum_{i' \in I} \pr^{|\phi_d\ra_{AB}} \lt( \neg \sum_{i \in I} [i]^A, [i']^B \rt)  = 0 \quad \Rightarrow \quad \sum_{i' \in I} \pr_\lam^{|\phi_d\ra_{AB}} \lt( \neg \sum_{i \in I} [i]^A, [i']^B  \rt) \iae 0.
\end{align}
Then, using $\pr(X) = \sum_{i \in I}\pr(X,Y_i) + \pr(X, \neg \bigvee_{i \in I} Y_i)$ for a set of mutually exclusive events $\{Y_i\}_{i \in I}$, we have for all $\hat{O}^A, \hat{O}^B$,
\begin{align}
\pr_\lam^{|\phi_d\ra_{AB}} \lt( \sum_{i \in I} [i]^A \rt)_{\hat{O}^A}
&= \sum_{i' \in I} \pr_\lam^{|\phi_d\ra_{AB}} \lt( \sum_{i \in I} [i]^A, [i']^B \rt)_{\hat{O}^A \otimes \hat{O}^B} + \pr_\lam^{|\phi_d\ra_{AB}} \lt( \sum_{i \in I} [i]^A, \neg \bigvee_{i' \in I} [i']^B \rt)_{\hat{O}^A \otimes \hat{O}^B} \notag\\
&\iae \sum_{i' \in I} \pr_\lam^{|\phi_d\ra_{AB}} \lt( \sum_{i \in I} [i]^A, [i']^B \rt)_{\hat{O}^A \otimes \hat{O}^B} + \sum_{i' \in I} \pr_\lam^{|\phi_d\ra_{AB}} \lt( \neg \sum_{i \in I} [i]^A, [i']^B \rt)_{\hat{O}^A \otimes \hat{O}^B} \notag\\
&= \sum_{i \in I} \pr_\lam^{|\phi_d\ra_{AB}} \lt([i]^B\rt)_{\hat{O}^B}.\label{PC0der}
\end{align}
By setting $I = \{i\}$ we have, for all $\hat{O}^A, \hat{O}^B$ and for all $i$,
\begin{align}
\pr_\lam^{|\phi_d\ra_{AB}} \lt( [i]^A \rt)_{\hat{O}^A} = \pr_\lam^{|\phi_d\ra_{AB}} \lt([i]^B\rt)_{\hat{O}^B},\label{PC0CI}
\end{align}
which means that both probabilities are context-independent, so the observables can be dropped. Therefore, from \eqref{PC0der} we arrive at
\begin{align}
\pr_\lam^{|\phi_d\ra_{AB}} \lt( \sum_{i \in I} [i]^A \rt) \iae \sum_{i \in I} \pr_\lam^{|\phi_d\ra_{AB}} \lt( [i]^B \rt).
\end{align}

To prove \eqref{PC3}, note that
\begin{align}
&\pr^{|\Psi_n\ra} \lt(\sum_{(i,j) \in I_l}\hat{E}_{(i,j),n,l},\hat{F}_{(i',j'),n,l}\rt) = \notag\\
&\la \Psi_{n} | \lt( U^{AA'A''}_{n,l} \otimes  U^{BB'B''}_{n,l}\rt)^{-1} \lt(\sum_{(i,j) \in I_l} [i,j]^{AA'} \otimes [i',j']^{BB'} \otimes \mathbb{I}^{A''B''} \rt) \lt( U^{AA'A''}_{n,l} \otimes  U^{BB'B''}_{n,l}\rt) | \Psi_{n} \ra,
\end{align}
where, by \eqref{DefU}, writing $\lfloor x \rfloor$ for the \emph{floor} of $x$, which is the greatest integer not greater than $x$,
\begin{align}
{\lt( U^{AA'A''}_{n,l} \otimes U^{BB'B''}_{n,l}\rt)|\Psi_n\ra} = \sum_{i=0}^{d-1} \sum_{k=0}^{n-1} c_{i,k} |i\ra_A |i\ra_B |k\;\mathrm{mod}\;m_{i,l}\ra_{A'}|k\;\mathrm{mod}\;m_{i,l}\ra_{B'} \lt| \lt\lfloor k/m_{i,l} \rt\rfloor \rt\ra_{A''} \lt| \lt\lfloor k/m_{i,l} \rt\rfloor \rt\ra_{B''}.
\end{align}
Using a similar strategy as above, we get
\begin{align}
\pr^{|\Psi_n\ra} \lt(\sum_{(i,j) \in I_l}\hat{E}_{(i,j),n,l},\neg\!\!\bigvee_{(i',j') \in I_l} \hat{F}_{(i',j'),n,l}\rt) &= 0 \notag\\
\Rightarrow \quad \pr_\lam^{|\Psi_n\ra} \lt(\sum_{(i,j) \in I_l}\hat{E}_{(i,j),n,l},\neg\!\!\bigvee_{(i',j') \in I_l} \hat{F}_{(i',j'),n,l}  \rt) &\iae 0; \notag\\
\sum_{(i',j') \in I_l} \pr^{|\Psi_n\ra} \lt(\neg\sum_{(i,j) \in I_l}\hat{E}_{(i,j),n,l}, \hat{F}_{(i',j'),n,l}\rt) &= 0 \notag\\
\Rightarrow \quad \sum_{(i',j') \in I_l} \pr_\lam^{|\Psi_n\ra} \lt(\neg\sum_{(i,j) \in I_l}\hat{E}_{(i,j),n,l}, \hat{F}_{(i',j'),n,l}  \rt) &\iae 0.
\end{align}
Then, for all $\hat{O}^{AA'A''}, \hat{O}^{BB'B''}$,
\begin{align}
&\pr_\lam^{|\Psi_n\ra} \lt(\sum_{(i,j) \in I_l} \hat{E}_{(i,j),n,l}\rt)_{\hat{O}^{AA'A''}}\notag\\
&= \sum_{(i',j') \in I_l} \pr_\lam^{|\Psi_n\ra} \lt(\sum_{(i,j) \in I_l}\hat{E}_{(i,j),n,l},\hat{F}_{(i',j'),n,l}  \rt)_{\hat{O}^{AA'A''} \otimes \hat{O}^{BB'B''}} \notag\\
&+ \pr_\lam^{|\Psi_n\ra} \lt(\sum_{(i,j) \in I_l}\hat{E}_{(i,j),n,l},\neg\!\!\bigvee_{(i',j') \in I_l} \hat{F}_{(i',j'),n,l}  \rt)_{\hat{O}^{AA'A''} \otimes \hat{O}^{BB'B''}} \notag\\
&\iae \sum_{(i',j') \in I_l} \pr_\lam^{|\Psi_n\ra} \lt(\sum_{(i,j) \in I_l}\hat{E}_{(i,j),n,l},\hat{F}_{(i',j'),n,l}  \rt)_{\hat{O}^{AA'A''} \otimes \hat{O}^{BB'B''}}\notag\\
&+ \sum_{(i',j') \in I_l} \pr^{|\Psi_n\ra} \lt(\neg\sum_{(i,j) \in I_l}\hat{E}_{(i,j),n,l}, \hat{F}_{(i',j'),n,l}  \rt)_{\hat{O}^{AA'A''} \otimes \hat{O}^{BB'B''}} \notag\\
&= \sum_{(i,j) \in I_l} \pr_\lam^{|\Psi_n\ra} \lt(\hat{F}_{(i,j),n,l}\rt)_{\hat{O}^{BB'B''}}.\label{PC3der}
\end{align}
By setting $I_l = (i,j)$, we get, for all $i,j$ and for all $\hat{O}^{AA'A''}, \hat{O}^{BB'B''}$,
\begin{align}
\pr_\lam^{|\Psi_n\ra} \lt( \hat{E}_{(i,j),n,l}\rt)_{\hat{O}^{AA'A''}} \iae \pr_\lam^{|\Psi_n\ra} \lt(\hat{F}_{(i,j),n,l}\rt)_{\hat{O}^{BB'B''}}.\label{PC3CI}
\end{align}
As in \eqref{PC0CI}, this means the probabilities are context-independent, and \eqref{PC3} follows from \eqref{PC3der}.

Similarly, to prove \eqref{PC1}, consider the probability
\begin{align}
&\pr^{|\Psi_n\ra} \lt(\hat{E}_{(i,j),n,l},[i']^B\rt) =\notag\\
&\la \Psi_n| \lt( U^{AA'A''}_{n,l} \otimes \mathbb{I}^{BB'B''}\rt)^{-1} \lt( [i,j]^{AA'} \otimes [i']^B \otimes \mathbb{I}^{A''B'B''} \rt) \lt( U^{AA'A''}_{n,l} \otimes \mathbb{I}^{BB'B''}\rt) |\Psi_n\ra,
\end{align}
where, by \eqref{DefU},
\begin{align}
\lt( U^{AA'A''}_{n,l} \otimes \mathbb{I}^{BB'B''}\rt)|\Psi_n\ra = \sum_{i=0}^{d-1} \sum_{k=0}^{n-1} c_{i,k} |i\ra_A |k\;\mathrm{mod}\;m_{i,l}\ra_{A'} \lt| \lt\lfloor {k}/{m_{i,l}} \rt\rfloor \rt\ra_{A''} \otimes |i\ra_B |0\ra_{B'} |k\ra_{B''}.
\end{align}
Again, using CompQuant, it follows that
\begin{align}
\sum_{j=0}^{m_{i,l}-1}\pr^{|\Psi_n\ra} \lt(\hat{E}_{(i,j),n,l}, \neg [i]^B\rt) = 0 &\quad \Rightarrow \quad \sum_{j=0}^{m_{i,l}-1}\pr_\lam^{|\Psi_n\ra} \lt(\hat{E}_{(i,j),n,l}, \neg [i]^B \rt) \iae 0; \notag\\
\pr^{|\Psi_n\ra} \lt(\neg\!\!\bigvee_{j=0}^{m_{i,l}-1}\hat{E}_{(i,j),n,l}, [i]^B\rt) = 0 &\quad \Rightarrow \quad \pr_\lam^{|\Psi_n\ra} \lt(\neg\!\!\bigvee_{j=0}^{m_{i,l}-1}\hat{E}_{(i,j),n,l}, [i]^B \rt) \iae 0.
\end{align}
Then, for all $\hat{O}^{AA'A''}, \hat{O}^B$,
\begin{align}
&\sum_{j=0}^{m_{i,l}-1}\pr_\lam^{|\Psi_n\ra} \lt(\hat{E}_{(i,j),n,l}\rt)_{\hat{O}^{AA'A''}}\notag\\
&=  \sum_{j=0}^{m_{i,l}-1}\pr_\lam^{|\Psi_n\ra} \lt(\hat{E}_{(i,j),n,l}, [i]^B  \rt)_{\hat{O}^{AA'A''} \otimes \hat{O}^B} + \sum_{j=0}^{m_{i,l}-1}\pr_\lam^{|\Psi_n\ra} \lt(\hat{E}_{(i,j),n,l}, \neg [i]^B  \rt)_{\hat{O}^{AA'A''} \otimes \hat{O}^B} \notag\\
&\iae \sum_{j=0}^{m_{i,l}-1}\pr_\lam^{|\Psi_n\ra} \lt(\hat{E}_{(i,j),n,l}, [i]^B  \rt)_{\hat{O}^{AA'A''} \otimes \hat{O}^B} + \pr_\lam^{|\Psi_n\ra} \lt(\neg\!\!\bigvee_{j=0}^{m_{i,l}-1}\hat{E}_{(i,j),n,l}, [i]^B \rt)_{\hat{O}^{AA'A''} \otimes \hat{O}^B} \notag\\
&= \pr_\lam^{|\Psi_n\ra} \lt([i]^B\rt)_{\hat{O}^B}.
\end{align}
Using the context-independence of the probability on the left-hand side of \eqref{PC3CI}, we arrive at \eqref{PC1}. Since the above derivation also holds when all $A$ and $B$ are interchanged, we immediately also get \eqref{PC2}.

\bibliographystyle{model5-names}
\bibliography{refs}

\end{document}